\documentclass{article}

\usepackage{microtype}
\usepackage{graphicx}
\usepackage{subfigure}
\usepackage{booktabs} %

\usepackage{hyperref}

\usepackage{amsmath}
\usepackage{bbm}
\usepackage{booktabs}
\usepackage{amsfonts}
\usepackage{selectp}
\usepackage{graphicx}
\usepackage{color}
\usepackage{enumitem}
\usepackage{amsfonts}
\usepackage{amssymb}
\usepackage{amsmath}
\usepackage{bm}
\usepackage{lipsum}
\usepackage{graphicx}
\usepackage[dvipsnames]{xcolor}
\usepackage{ragged2e}
\usepackage{multirow}
\usepackage{pifont}

\usepackage{multicol}
\usepackage{caption}
\usepackage{subcaption}
\usepackage{mathtools}
\usepackage{microtype}
\usepackage{changepage}
\usepackage{graphicx}

\usepackage{pifont}
\usepackage{amsmath}
\usepackage{amssymb}
\usepackage{mathtools}
\usepackage{amsthm, bm}
\newcommand{\defeq}{\vcentcolon=}
\theoremstyle{plain}

\theoremstyle{definition}

\theoremstyle{remark}

\usepackage[colorinlistoftodos,textwidth=3.0cm,textsize=tiny]{todonotes}

\newcommand{\acrolong}{\textbf{D}iffusion \textbf{I}nference-\textbf{T}ime $\boldsymbol{T}$-\textbf{O}ptimization}

\newcommand{\acro}{DITTO}
\definecolor{chng}{rgb}{0, 0., 0.}

\newcommand{\eref}[1]{(\ref{#1})}
\newcommand{\aref}[1]{Appendix~\ref{#1}}
\newcommand{\algref}[1]{Algorithm~\ref{#1}}
\newcommand{\sref}[1]{Section~\ref{#1}}
\newcommand{\fref}[1]{Fig.~\ref{#1}}
\newcommand{\tref}[1]{Table~\ref{#1}}
\usepackage{array}
\usepackage{multirow}
\usepackage{booktabs}
\usepackage{caption}
\usepackage{float}
\usepackage{longtable}
\usepackage{algorithm}
\usepackage{algorithmic}
\usepackage{graphicx}
\definecolor{good}{rgb}{0.11, 0.77, 0.11}
\definecolor{bad}{rgb}{0.77, 0.11, 0.11}
\definecolor{ditto}{RGB}{93,84,164}
\usepackage{hyperref}
\usepackage{url}

\usepackage[accepted]{icml2024}

\usepackage{amsmath}
\usepackage{amssymb}
\usepackage{mathtools}
\usepackage{amsthm}

\usepackage[capitalize,noabbrev]{cleveref}

\theoremstyle{plain}
\theoremstyle{definition}
\theoremstyle{remark}

\usepackage[textsize=tiny]{todonotes}

\icmltitlerunning{\acro{}: \acrolong{} for Music Generation}

\begin{document}

\twocolumn[
\icmltitle{DITTO: Diffusion Inference-Time \texorpdfstring{$\bm{T}$} --Optimization for Music Generation}

\icmlsetsymbol{ack}{*}

\begin{icmlauthorlist}
\icmlauthor{Zachary Novack}{yyy,comp,ack}
\icmlauthor{Julian McAuley}{yyy}
\icmlauthor{Taylor Berg-Kirkpatrick}{yyy}
\icmlauthor{Nicholas J. Bryan}{comp}
\end{icmlauthorlist}

\icmlaffiliation{yyy}{University of California -- San Diego}
\icmlaffiliation{comp}{Adobe Research}

\icmlcorrespondingauthor{Zachary Novack}{znovack@ucsd.edu}
\icmlcorrespondingauthor{Nicholas J. Bryan}{njb@ieee.org}

\vskip 0.3in
]

\icmlkeywords{Machine Learning, ICML}
\printAffiliationsAndNotice{$^*$Work done during an internship at Adobe Research.}  

\begin{abstract}
We propose \acrolong{} (\textbf{\acro{}}), a general-purpose framework for controlling pre-trained 
text-to-music 
diffusion models at inference-time via optimizing initial noise latents. Our method can be used to optimize through any differentiable feature matching loss to achieve a target (stylized) output and leverages gradient checkpointing for memory efficiency. 
We demonstrate a surprisingly wide-range of applications for music generation including inpainting, outpainting, and looping as well as intensity, melody, and musical structure control -- all 
without ever fine-tuning the underlying model. When we compare our approach against related training, guidance, and optimization-based methods, we find \acro{} achieves state-of-the-art performance on nearly all tasks,
including outperforming comparable approaches on controllability, audio quality, and computational efficiency, thus opening the door for high-quality, flexible, training-free control of diffusion models. Sound examples can be found at \url{https://ditto-music.github.io/web/}.
\end{abstract}

\begin{figure*}
    \centering
\includegraphics[width=\textwidth]{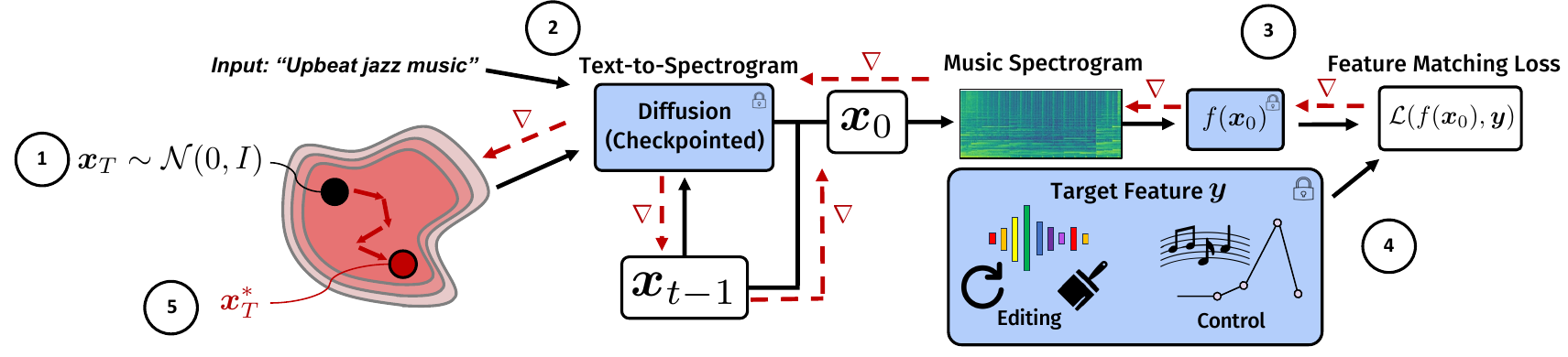}
    \caption{We propose \textbf{\acro{}}, or \acrolong{},
    a general-purpose framework to control pre-trained diffusion models at inference-time. 1) We sample an initial noise latent $\bm{x}_T$; 2) run diffusion sampling to generate a music spectrogram $\bm{x}_0$; 3) extract features from the generated content; 4) input a target control signal; and 5) optimize the initial noise latent to fit any differentiable loss.}
    \label{fig:ditto}
\end{figure*}

\section{Introduction}

Large-scale diffusion models~\citep{ho2020denoising} have emerged as a
leading paradigm for generative media, with strong results in diverse modalities such as text-to-image (TTI) generation~\citep{rombach2022high, karras2022elucidating, chen2023importance}, video generation~\citep{ho2022video, Gupta2023PhotorealisticVG}, and 3D object generation~\citep{Watson2022NovelVS, poole2022dreamfusion}. 
Recently, there has been growing work in applying image-domain methods to 
audio
by treating the frequency-domain spectrograms of audio as images, producing promising results in general text-to-audio (TTA) generation~\citep{liu2023audioldm, liu2023audioldm2, huang2023make} and text-to-music (TTM) generation~\citep{hawthorne2022multi, forsgren2022riffusion, chen2023musicldm, huang2023noise2music, schneider2023mo}. 
These methods operate via pixel or latent diffusion~\cite{rombach2022high} over spectrograms with genre, mood, and/or keywords control articulated via text prompts. 

However, these text-conditioned approaches
typically only offer high-level 
control
(e.g.~style), motivating further work.
Current attempts to add more precise 
control 
(e.g.~time-varying conditions)
for TTM diffusion models are promising yet
present their own tradeoffs.
Finetuning-based methods like ControlNet~\citep{Wu2023MusicCM, saharia2022palette, zhang2023adding}
require 
large-scale
supervised
training 
with labeled examples for each new control modality.
Inference-time methods that guide the diffusion sampling process, on the other hand, struggle to achieve fine-grained expressivity due to relying on approximations of the model outputs during sampling~\citep{Levy2023ControllableMP, yu2023freedom}.

In order to achieve an expressive control paradigm for TTM diffusion models that requires no supervised training and can accept arbitrary control signals at inference-time,
we propose \textbf{\acro{}}: \acrolong{}. \acro{} optimizes the initial noise latents $\bm{x}_T$ with respect to an \emph{arbitrary} differentiable feature matching loss across any 
diffusion sampling process 
to control the model outputs,
and ensures efficient memory use via gradient checkpointing~\citep{chen2016training}.
Despite generally being considered 
to encode little information \citep{song2020denoising, preechakul2022diffusion}, 
we show the power and precision the initial noise latents have to 
control the diffusion process for a wide-variety of applications in music creation, enabling musically-salient feature control and 
high-quality 
audio editing.
Compared to previous optimization-based works from outside the audio domain~\citep{Wallace2023EndtoEndDL},
\acro{} achieves SOTA control
while also being 2x as time and memory efficient.
Overall, our contributions are:
\vspace{-0.25cm}
\begin{itemize}
\item \acro{}: a 
novel,
training-free framework for controlling pre-trained TTM diffusion models 
that optimizes the initial noise latents 
to control the model outputs.
\item We 
leverage
gradient checkpointing for memory efficiency without 
compromising the sampling process.
\item Application of 
\acro{}
to  
multiple fine-grained time-dependent tasks, including 
audio-domain 
inpainting, outpainting, melody control, intensity control, and the newly proposed looping and musical structure control.
\item Evaluation showing our approach outperforms MultiDiffusion~\citep{bar2023multidiffusion}, FreeDoM~\citep{yu2023freedom}, Guidance Gradients~\cite{Levy2023ControllableMP}, Music ControlNet~\citep{Wu2023MusicCM}, and the comparable 
optimization method DOODL~\cite{Wallace2023EndtoEndDL}, while 
being 2x faster and using 
half the
memory.
\end{itemize}

\section{Related Work}

\subsection{Music Generation Overview}
Early works on generative music focused on \emph{symbolic} generation 
\citep{dong2018musegan, chen2020music, dai2021controllable}. Recently, \emph{audio}-domain music generation has become popular due to advances in language models (LMs) like MusicLM~\citep{agostinelli2023musiclm} and diffusion models like AudioLDM~\citep{liu2023audioldm, liu2023audioldm2}.  
LM-based approaches typically operate over discrete compressed audio tokens~\citep{zeghidour2021soundstream, kumar2023high}, generating audio 
either
autoregressively~\citep{borsos2023audiolm, agostinelli2023musiclm, copet2023simple} or 
non-autoregressively
\citep{garcia2023vampnet, Borsos2023SoundStormEP}, and convert generated tokens back to audio directly. Diffusion-based approaches, on the other hand, typically operate by generating 2D frequency-domain representations of audio or \emph{spectrograms} that are decoded into audio via 
a vocoder~\citep{forsgren2022riffusion, liu2023audioldm, liu2023audioldm2, schneider2023mo}.

\subsection{Diffusion Models with Text Control}
Text is currently the most popular control medium for diffusion models. 
Here,
text captions are encoded into embeddings and injected into a generative model during training via cross attention, additive modulation, or similar as found in 
Stable Diffusion~\citep{rombach2022high} or Imagen \citep{saharia2022photorealistic}. Despite its popularity, global caption-based text conditioning lacks fine-grained control \citep{zhang2023adding}, motivating alternatives
and the present work.

\subsection{Alternative Train-time Control Methods}
It is common to
fine-tune existing text-conditioned diffusion models with additional inputs when adding advanced control. ControlNet-type models~\citep{zhang2023adding, zhao2023uni} 
use large sets of paired data to fine-tune 
TTI
diffusion models by adding 
control adapters for specific predefined controls such as edge detection or pose estimation. 
To reduce training demands, 
a number of works fine-tune pre-trained models on a small number of examples~\citep{ruiz2023dreambooth, choi2023custom, gal2022textual, kawar2023imagic}. Others have explored using external reward models for fine-tuning, through direct fine-tuning~\citep{Clark2023DirectlyFD, Prabhudesai2023AligningTD} or reinforcement learning~\citep{black2023training}. Such approaches, however, still require an expensive training process and the control mechanism cannot be modified after training. For music, only a ControlNet-style approach has been taken~\cite{Wu2023MusicCM}. In contrast, \acro{} requires no large-scale training and can accept any differentiable
control at inference-time.

\subsection{Inference-time Guidance-based Control}
To avoid 
large-scale model fine-tuning, inference-time control methods 
have become increasingly popular. 
Early approaches 
include prompt-to-prompt image editing~\cite{hertz2022prompt} and MultiDiffusion~\cite{bar2023multidiffusion}, which enable localized object
editing and
in/outpainting 
by fusing multiple masked diffusion paths together. Such methods rely on control targets that can be localized to specific pixel regions 
of an image 
and are less applicable for audio spectrograms which have indirect pixel correspondences across frequency and multiple overlapping sources at once. 

We also note the class of
guidance-based methods~\citep{dhariwal2021diffusion, Chung2022DiffusionPS, Levy2023ControllableMP, yu2023freedom}, 
which introduce updates at each sampling step to steer the generation process 
via the gradient of a pre-trained classifier $\nabla_{x_t}\mathcal{L}_\phi(x_t)$.
These approaches 
generally
require 
an approximation of model outputs during sampling, which 
are inaccurate at high noise levels and thus 
limit fine-grained expressivity. 
For music, guidance-based methods have only been explored in~\citet{Levy2023ControllableMP}. In contrast, \acro{} calculates gradients with respect to the initial noise on the \emph{real} model outputs through sampling, allowing accurate gradients to influence the entire generation process.

\subsection{Inference-time Optimization-based Control}
Recent work has shown optimization
through diffusion sampling is possible if GPU memory is 
correctly managed.
Direct optimization of diffusion latents (DOODL)~\citep{Wallace2023EndtoEndDL} leverages the recently proposed EDICT 
sampler~\citep{wallace2023edict}, which uses affine coupling layers (ACLs)~\cite{dinh2014nice, dinh2016density} to form a fully invertible sampling process, and backpropagates through EDICT to optimize initial 
noise latents for improving
high-level features like CLIP guidance 
and aesthetic improvement
in images. 
DOODL,
in contrast to our approach,
struggles on fine-grained control signals~\citep{Wallace2023EndtoEndDL} and 
has 
multiple
downsides due to its reliance on EDICT including 1) it is 
restricted to only invertible 
sampling algorithms;
2) it requires double the model evaluations for both forward and reverse sampling that increase latency and memory use; and 3) it can suffer from stability issues and reward hacking due to divergence between the ACL diffusion chains.

\citet{karunratanakul2023optimizing}
proposed backpropagating through 
sampling 
for human motion generation
(i.e.~short sequences of 
joint positions). This work leverages numerous domain-specific modifications 
to reduce memory usage, such as using a small (i.e.~$<$18M parameters) transformer encoder-only architecture, very few sampling steps, long optimization time, and purely unconditional generation. Thus, this approach is not applicable to 
more standard 
generative tasks with higher memory demands like text-to-image/audio/music, 
while \acro{}
circumvents any
restrictions on the model architecture or sampler.

\section{Diffusion Inference-Time $\bm{T}$-Optimization}

\subsection{Diffusion Background}
Denoising~diffusion~probabilistic~models~(DDPMs)~\citep{sohl2015deep,ho2020denoising} or diffusion models are defined by a forward and reverse random Markov process. The forward process takes clean data and iteratively corrupts it with noise to train a neural network $\bm{\epsilon}_\theta$. The network $\bm{\epsilon}_\theta$ typically inputs (noisy) data $\bm{x}_{t}$, the diffusion step  $t$, and (text) conditioning information $\bm{c}$. The reverse process takes random noise $\bm{x}_{T} \sim \mathcal{N}(0,I)$ and iteratively refines it with the learned network to generate new data $\bm{x}_{0}$ over 
$T$ 
time steps (e.g., $1000$)
via the sampling process,
\begin{equation}
    \bm{x}_{t-1} = \frac{1}{\sqrt{\alpha_t}} \Big ( \bm{x}_{t} - \frac{1-\alpha_t}{\sqrt{1-\bar{\alpha}_t} } \bm{\epsilon}_\theta(\bm{x}_{t}, t, \bm{c}) \Big) + \sigma_t  \bm{\epsilon}, \label{eq:DDPM}
\end{equation}
where $\bm{\epsilon} \sim \mathcal{N}(0,I)$, $\alpha_0 \defeq 1$, $\alpha_t$ and $\bar{\alpha}_t$ 
define the noise schedule, $\sigma_t$ is the sampling standard deviation. 
To reduce sampling time, Denoising Diffusion Implicit Model (DDIM) sampling~\cite{song2020denoising}
uses an alternative optimization objective that yields a faster sampling process
(e.g., $20-50$ steps) that can be deterministic. Broadly, we can denote any sampling algorithm with the notation $\bm{x}_{t-1} = \texttt{Sampler}(\bm{\epsilon}_\theta, \bm{x}_t, t, \bm{c})$.

To improve text conditioning, classifier-free guidance (CFG) can be used to blend conditional and unconditional generation outputs~\cite{ho2022classifier}.
 When training 
with CFG, conditioning 
is randomly set to a null value
a fraction of the time. 
During inference, the diffusion model output $\bm{\epsilon}_\theta(\bm{x}_{t}, t, \bm{c})$ 
is linearly combined with $\bm{\epsilon}_\theta(\bm{x}_{t}, t, \bm{c}_\emptyset)$ using the CFG scale $w$,
where 
$\bm{c}_\emptyset$ are null embeddings. Note, CFG during inference doubles the forward passes of $\bm{\epsilon}_\theta$.
For a diffusion model review, see~\aref{appendix:diffusion}. 

\subsection{Problem Formulation}

Instead of trying to control diffusion models by using 
expensive supervised training or 
inexact inference-time guidance-based methods, 
we alternatively formulate the control task
as an \emph{optimization} problem. 
Notably, we can denote the output of the model after running the sampler for a total of $T$ sampling steps as $\bm{x}_0 = \texttt{Sampler}_T(\bm{\epsilon}_\theta, \bm{x}_T, \bm{c})$, showing that the final output is a function of the \emph{initial} noise latents $\bm{x}_T \sim~\mathcal{N}(0, I)$.

While $\bm{x}_T$ is normally just considered to be a random seed, we can instead treat the initial noise latents as a free parameter to be optimized at inference-time. In particular, we define a target feature extractor $f(\cdot)$, which only needs to be differentiable, and some corresponding loss function $\mathcal{L}$ to measure how well the model output's particular feature matches a target control $\bm{y}$. With this, we can then directly optimize $\bm{x}_T$ \emph{through} the sampling process such that the model output $\bm{x}_0$ follows the target control.
Formally,
\begin{align}
    \boldsymbol{x}_T^* &= \arg \min_{\boldsymbol{x}_T} \mathcal{L}\left(f(\boldsymbol{x}_0), \boldsymbol{y}\right) \label{eq:problem}\\
    \bm{x}_0 &= \texttt{Sampler}_T(\bm{\epsilon}_\theta, \bm{x}_T, \bm{c})\label{eq:sampler}
\end{align}
By framing the control task as an arbitrary feature-matching optimization on the initial noise latents, we are able to incorporate a diverse range of control tasks by constructing $f(\cdot)$ and $\mathcal{L}$ accordingly, such as letting $f$ extract the intensity curve of the music and $\mathcal{L}$ being the squared $\ell_2$ distance to some target intensity (see Sec.~\ref{sec:apps} for more details). This procedure requires no training (as only $\bm{x}_T$ is optimized rather than model weights) and uses \emph{exact} control gradients (as $f(\cdot)$ is only called on the real output).

Solving~\eref{eq:problem} using backpropagation, however, is typically intractable due to extreme memory requirements. Namely, the diffusion sampling process is recursive
by design 
and standard automatic differentiation packages customarily require storing all intermediate results for each of $T$ recurrent calls to $\bm{\epsilon}_\theta$ within the sampler ($2 T$ sets of activations per step when CFG is used). Thus, even 2-3 sampling steps can cause memory errors with standard U-Net diffusion architectures.

\begin{algorithm}[t]
    \caption{ 
    \acrolong{} (\acro{})}
    \label{alg:ditto}
    \begin{algorithmic}[1]
    \INPUT: $\bm{\epsilon}_\theta$, \texttt{Sampler}, sampling steps $T$, feature extractor $f$, loss $\mathcal{L}$, target feature $\boldsymbol{y}$, starting latent $\boldsymbol{x}_T$, text conditioning $\bm{c}$, optimization steps $K$, optimizer $g$.
        \vspace{3pt}
        \STATE // Run optimization
        \FOR{$i=1$ to $K$}
            \STATE // Initialize noise latents
            \STATE $\boldsymbol{x}_t \leftarrow \boldsymbol{x}_T$ %
            \STATE // Diffusion sampling w/grad checkpointing per step
            \FOR{$t=T$ to $1$}
                \STATE $\boldsymbol{x}_{t-1} = \texttt{Checkpoint}( \texttt{Sampler}, \bm{\epsilon}_\theta, \boldsymbol{x}_t, t, \bm{c})$ %
            \ENDFOR
            \STATE // Extract features from generated output
            \STATE $\hat{\boldsymbol{y}} = f(\boldsymbol{x}_0)$ %

            \STATE // Compute the loss and backprop
            \STATE $\boldsymbol{x}_T \leftarrow \boldsymbol{x}_T - g(\nabla_{\boldsymbol{x}_T}\mathcal{L}(\hat{\boldsymbol{y}}, \boldsymbol{y}))$ %
        \ENDFOR

    \OUTPUT: $\boldsymbol{x}_0$
    
\end{algorithmic}
\end{algorithm}

\begin{figure}
    \centering
\includegraphics[width=0.5\textwidth]{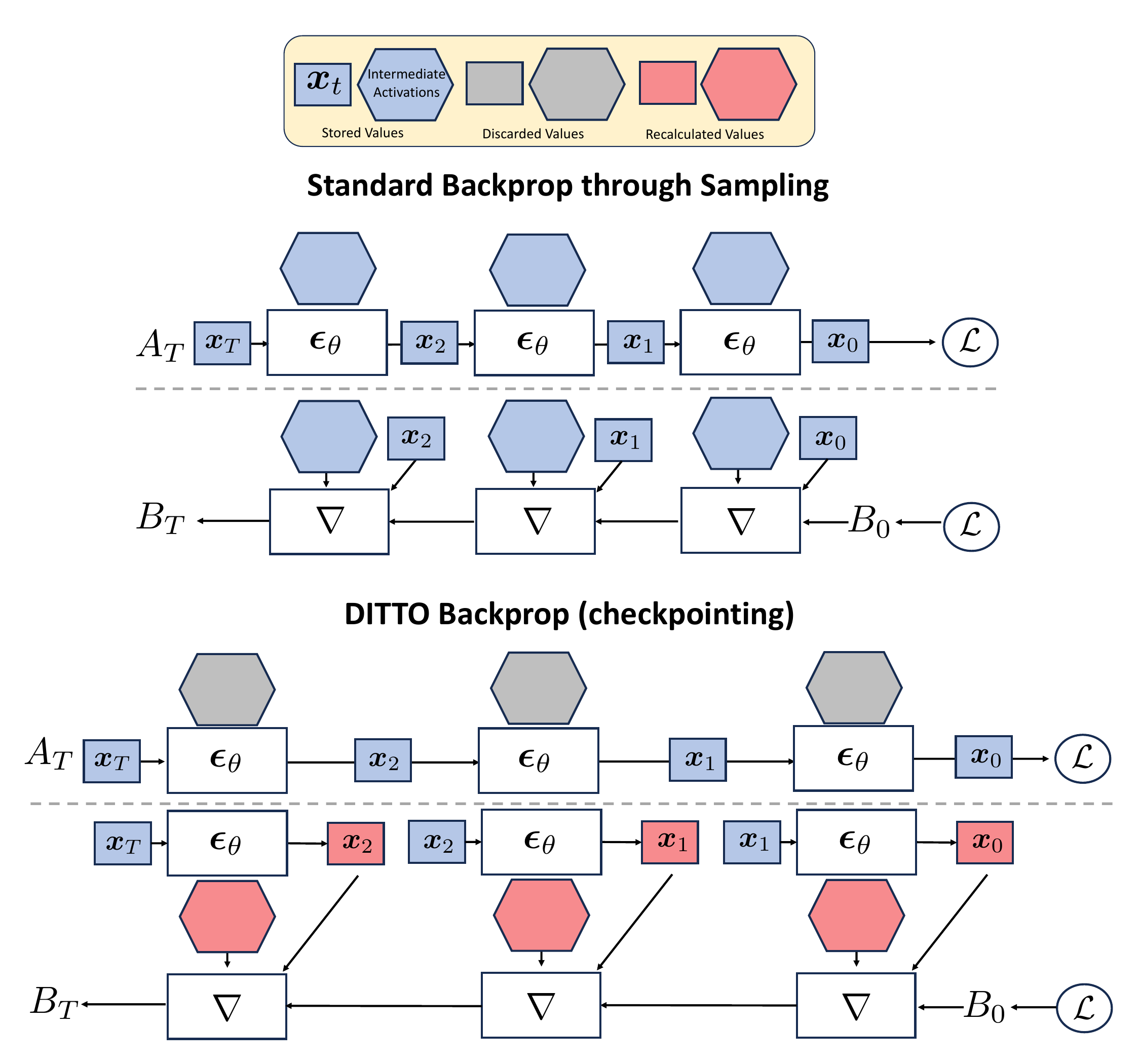}
    \caption{Different memory setups for backpropagation through sampling. Normally, all intermediate activations are stored in memory, which is intractable for modern diffusion models.
    In \acro{}, gradient checkpointing allows us to achieve efficient memory usage with only 2x the number of model calls to preserve fast runtime. }
    \label{fig:checkp}
\end{figure}
\subsection{Diffusion with Gradient Checkpointing}

To circumvent large memory use during optimization, we use gradient 
checkpointing~\citep{chen2016training}. 
The core idea is to discard intermediate activation values stored during the forward pass of backpropagation that inflict high memory use 
and recalculate them during the backward pass when needed from cached inputs. 
We use gradient checkpointing on each model call during sampling,
as the memory required to store the intermediate noisy diffusion tensors and conditioning information is 
minute compared to  the intermediate activations of a typical diffusion model (e.g., cross-attention activation maps within a large UNet). 
Our memory cost to optimize~\eref{eq:problem} with sampler-step checkpointing is 1) the memory needed to run backpropagation on one diffusion model call $\bm{\epsilon}_\theta$ plus 2) the cost to store the $T$ intermediate noisy diffusion tensors $\bm{x}_t \forall t=0,...,T$ and conditioning  $\bm{c}$. 
While we pay for the memory reduction with
an additional forward pass 
per time step
(as shown in~\fref{fig:checkp}),
this straightforward trick allows \acro{} to maintain efficiency without changing any part of the sampling algorithm. 

In contrast to our approach, DOODL explored gradient checkpointing via the MemCNN library~\cite{vandeLeemput2019MemCNN}. However, their use of the EDICT sampler
\emph{doubles} the memory and runtime cost compared to our method (see Appendix~\ref{sec:edict_doodl}) and adds instability to the sampling process due EDICT's dual-chain sampling (see \sref{sec:perfcomp}).

\subsection{Complete Algorithm}
Psuedo-code for our \acro{} algorithm is shown in~\algref{alg:ditto}. 
We define $\texttt{Checkpoint}$ to be a gradient checkpointing function that 1) inputs and stores a callable differentiable network (i.e., the sampler) and any input arguments to the network, 2) overrides the default activation caching behavior of the network to turn off activation caching during the forward pass of backpropagation and 3) recomputes activations when needed in the backward pass. Note that in practice, we typically use a small subsequence of sampling steps (e.g. 20) spanning from $\bm{x}_T$ to $\bm{x}_0$.
 
\section{Applications and Control Frameworks}\label{sec:apps}

\begin{figure*}[ht!]
    \centering
    \includegraphics[width=\textwidth]{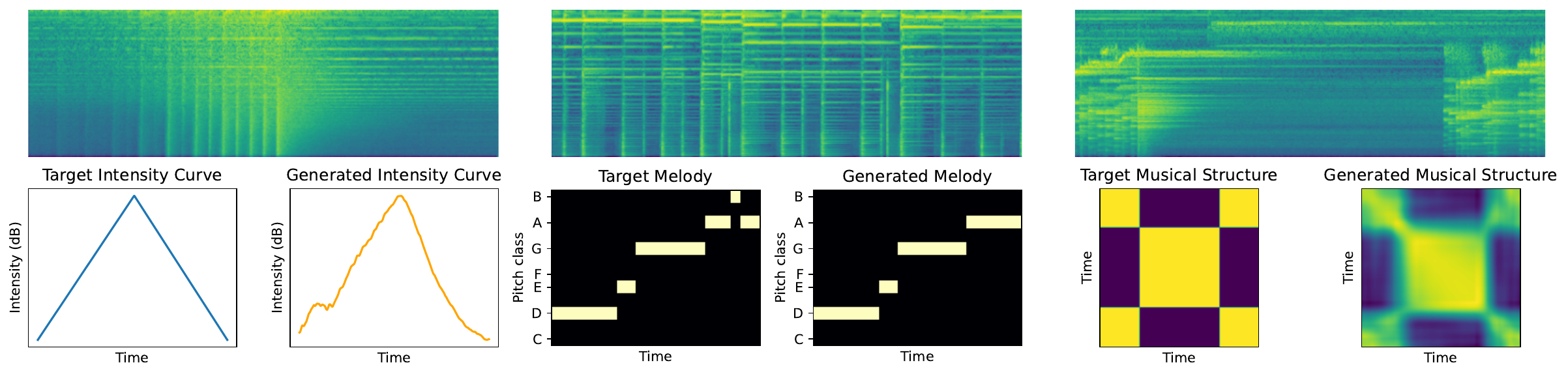}
    \vspace{-8mm}
    \caption{Examples of \acro{}'s use for creative control, including intensity (left), melody (middle), and structure (right), with target controls and final features displayed below each spectrogram. All results are achieved without additional training or fine-tuning.}
    \label{fig:contr}
\end{figure*}

We apply our flexible paradigm to a range of applications\footnote{We leave the rhythm control from \citet{Wu2023MusicCM} for future work, as their RNN beat detector would trigger an exceedingly long backpropagation graph when using \acro{}.} by parameterizing each control framework (i.e.~$f$ and $\mathcal{L}$) to directly target musically-salient features, 
allowing for
outpainting, inpainting, looping,
intensity control, melody control, and musical structure control, 
where musical structure and looping have been unexplored for TTM diffusion models. 
These constitute both reference-based (i.e.~using existing audio) and reference-free (generation from scratch, as shown in~\fref{fig:contr}) control operations.
Our goal here is to 
display the 
expressive
controllability 
that initial noise latents have over the diffusion process.

\textbf{Outpainting} -- Outpainting is the task of extending the length of 
existing audio
and is critical for 
audio editing as well as generating long-duration music content using diffusion models.  
Past outpainting methods include MultiDiffusion~\cite{bar2023multidiffusion} and Guidance Gradients~\cite{Levy2023ControllableMP} 
which struggle to maintain long-form coherence and local smoothing. We perform outpainting by 1) taking an existing reference audio signal $\bm{x}_{\text{ref}}$; 2) defining an overlap region $o$ in seconds at the end of the reference; 3) using \acro{} to create new content that matches the overlap region 
at the \emph{beginning} of the new generation;
and 4) stitching the reference and newly generated content together. More formally, 
we define $\mathbf{M}_{\text{ref}}$ and $\mathbf{M}_{\text{gen}}$ as binary masks that specify the location of the overlap region in the reference and generated content respectively, $f(\bm{x}_0)\coloneqq \mathbf{M}_{\text{gen}} \odot \bm{x}_0$, $\bm{y} = \mathbf{M}_{\text{ref}} \odot \bm{x}_{\text{ref}}$, and $\mathcal{L} \propto ||f(\bm{x}_0)- \bm{y}||_2^2$.

\textbf{Inpainting} -- Inpainting 
is the task of replacing an interior region of real or previously generated content and is essential for audio editing and music remixing.
Past work on inpainting
has been explored in the image- and audio-domain to variable success \citep{ Chung2022DiffusionPS, Levy2023ControllableMP}.  We use \acro{} to perform inpainting 
similar
to outpainting,
with the only modification being $\mathbf{M}_{\text{ref}} = \mathbf{M}_{\text{gen}}$ denote \emph{two} overlap regions (on each side of the spectrogram) to use as context for inpainting the gap in between.

\textbf{Looping} -- Looping is the task of generating content that repeats in a circular pattern, 
creating 
repeatable music fragments to form the basis of a larger composition. 
For looping, we use \acro{} similar to outpainting, but when we define $\mathbf{M}_{\text{ref}}$ and $\mathbf{M}_{\text{gen}}$, we specify two overlapping edge regions of the output (similar to inpainting) but corresponding to \emph{opposite} sides of the outputs (similar to outpainting), such that the extended region seamlessly transitions back to the beginning of the reference clip.  To our knowledge, we are the first to imbue TTM diffusion models with looping control.

\textbf{Intensity Control} -- Musical intensity control is the task of adjusting the dynamic contrast of generated music across time.  
We follow the intensity control protocol from
Music ControlNet (see~\citet{Wu2023MusicCM} for more details), 
which employs a training-time method to generate music that follows a smoothed, decibel (dB) volume curve. In our case, we use \acro{} in a similar fashion, albeit without the need for large-scale fine-tuning, by setting $f(\bm{x}_0) \coloneqq \bm{w} \ast 20\log_{10}(\texttt{RMS}(\mathbf{V}(\bm{x}_0)))$, where $\bm{w}$ are the smoothing coefficients used in Music ControlNet, $\ast$ is a convolution operator, $\texttt{RMS}$ is the Root Mean Squared energy of the audio, $\bm{y}$ is a given dB-scale target curve,  $\mathcal{L} \propto ||f(\bm{x}_0)- \bm{y}||_2^2$, and $\mathbf{V}$ is our 
vocoder~\cite{lee2022bigvgan, Zhu2024MusicHiFiFH} that translates spectrograms to the audio domain. Here, we backpropagate through our vocoder as well. Notably, under this parameterization intensity control does not only control the loudness of the generated audio but also the harmonic and rhythmic density of the music (which is correlated with RMS energy).

\textbf{Melody Control} -- Musical melody control is the task of controlling prominent musical tones over time and allows creators to generate accompaniment music to existing melodies. Following recent work~\citep{copet2023simple, Wu2023MusicCM}, the approx. melody of a recording can be extracted by computing the smoothed energy level of the 12-pitch classes over time via a highpass chromagram function $\mathbf{C}(\cdot)$~\cite{muller2015fundamentals}. Given this, we use \acro{} with $f(\bm{x}_0) = \log(\mathbf{C}(\mathbf{V}(\bm{x}_0)))$, a target melody  $\bm{y} \in \{1, \dots, 12\}^{N \times 1}$, the spectrogram length $N$, and $\mathcal{L} = \text{NLLLoss}(f(\bm{x}_0), \bm{y})$ or the negative log likelihood loss. See~\citet{Wu2023MusicCM} for further implementation details.

\textbf{Musical Structure Control} -- We define musical structure control as the task of controlling the high-level musical form of generated music over time. To model musical form, we follow musical structure analysis work~\cite{mcfee2014analyzing} that, in the simplest case, measures structure via computing a self-similarity (SS) matrix of local timbre features where timbre is ``everything about a sound which is neither loudness nor pitch''~\cite{erickson1975sound}. Thus, we use \acro{} for musical structure control by setting $\mathbf{y}$ to be a known, target SS matrix, $f(\bm{x}_0) = \mathbf{T}(\bm{x}_0) \mathbf{T}(\bm{x}_0)^\top$, $\mathbf{T}(\cdot)$ to be a timbre extraction function, and $\mathcal{L} \propto ||f(\bm{x}_0) - \bm{y}||_2^2$. 
Specifically, we use the Mel-Frequency Cepstrum Coefficients (MFCCs)~\citep{mcfee2010learning}, omitting the first coefficient and normalized across the time axis, as the timbre extraction function, 
and then smooth the SS matrix via a 2D Savitzky-Golay filter in order to not penalize slight variations in intra-phrase similarity. Such target SS matrices can take the form of an ``ABBA" pattern (as shown in~\fref{fig:contr}) for instance.
To our knowledge, we are the first to imbue TTM diffusion models with structure control.

\textbf{Other Applications} -- Besides the applications described above, \acro{} can be used for numerous new extensions previously unexplored in TTM generation which we describe in the Appendix, such as correlation-based intensity control (\ref{sec:corr_rms}), real-audio inversion (\ref{sec:inversion}), reference-free looping (\ref{sec:norefloop}), 
musical structure transfer (\ref{sec:fgstruct}),
other sampling methods (\ref{sec:alt_samp}), multi-feature optimization (\ref{sec:multi_obj}), and reusing optimized latents for fast inference (\ref{sec:reuse}). 

\section{Experimental Design}

\subsection{\acro{} Setup}

We use Adam~\citep{kingma2014adam} as our optimizer for \acro{}, with a learning rate of $5\times10^{-3}$ (as higher leads to stability issues). We use DDIM \citep{song2020denoising} sampling with $20$ steps and dynamic thresholding \citep{saharia2022photorealistic} for all experiments. 
No optimizer hyperparameters were changed across application besides the max number of optimization steps, which were doubled from 70 to 150 for the melody and structure tasks. 

\subsection{Datasets}
We train our models on a dataset of ${\approx}1800$ hours of licensed instrumental music with genre, mood, and tempo tags.
Our dataset does not have free-form text description,
so we use class-conditional text control of global musical style, as done in JukeBox~\cite{dhariwal2020jukebox}. For melody control references, we synthesize recordings from a 380-sample public-domain subset of the \textbf{Wikifonia Lead-Sheet Dataset}~\cite{simonetta2018symbolic}. Like in \citet{Wu2023MusicCM}, we construct a small set of handcrafted intensity curves and musical structure matrices (e.g.~a smooth crescendo and ``ABA" form) for intensity and structure control (see Appendix~\ref{sec:multi_obj} for more examples). For evaluation only, we also use the \textbf{MusicCaps Dataset}~\cite{agostinelli2023musiclm} with around 5K 10-second clips with text descriptions.

\subsection{Evaluation Metrics}
We use Frechet Audio Distance (FAD) with the CLAP music  \citep{wu2023large} backbone (as the default VGGish backbone is documented to poorly correlate with human perception \cite{Gui2023AdaptingFA}), which measures the distance between the distribution of embeddings from a set of baseline recordings and that from generated recordings~\citep{kilgour2018fr}. FAD metrics are calculated using MusicCaps as the reference distribution against 2.5K model generations for all experiments.
For reference-free targets, we also use the CLAP score \citep{wu2023large}, which measures the overall alignment between the text caption and the output audio; note that as our model is only \emph{tag}-conditioned, we convert each tag set into a caption using the template \emph{``A [genre] [mood] song at [BPM] beats per minute"}. Additionally, for the intensity and musical structure control, we report the average loss $\mathcal{L}$ across the generated outputs (i.e.~the final feature matching distance), and report overall accuracy for melody control, since it is framed as a classification task.

\subsection{Baselines}
We benchmark against a wide-range of methods including:
\begin{itemize}
    \item Naïve Masking: Here, after a DDIM-step we apply the update $\boldsymbol{x}_{t-1} = \mathbf{M}_{\text{ref}} \odot \mathcal{N}(\sqrt{\bar{\alpha}_t}\bm{x}_{\text{ref}}, (1-\bar{\alpha}_t) \bm{I}) +\mathbf{M}_{\text{gen}} \odot \boldsymbol{x}_{t-1}$ (i.e. setting the overlap region directly to the reference image at the appropriate noise level).
    \item MultiDiffusion \citep{bar2023multidiffusion}: This case is similar to the naïve approach, but instead \emph{averages} the noisy outputs in the overlapping region instead of using a hard mask. We can additionally stop this averaging operation at certain points of the sampling process (such as half way through) and let the model sample without guiding the process; we denote the former approach as MD and the latter as MD-50 for brevity.
    \item FreeDoM \citep{yu2023freedom}: FreeDoM is a guidance-based method, where
    we perform an additional update during sampling $\bm{x}_t = \bm{x}_t - \eta_t \nabla_{\boldsymbol{x}_t}\mathcal{L}(f(\hat{\boldsymbol{x}}_0(\boldsymbol{x}_t)), \boldsymbol{y})$, where $\hat{\boldsymbol{x}}_0(\boldsymbol{x}_t)$ denotes the first term in Eq.~\ref{eq:DDIM}. $\eta_t$ is a time-dependent learning rate that is a function of the overall gradient norm.
    \item Guidance Gradients (GG) \citep{Levy2023ControllableMP}: GG takes the update equation from FreeDoM and makes two small modifications. Namely, $\eta_t$ is fixed throughout sampling, and GG includes an additional data consistency step when the feature extractor $f(\cdot)$ is fully linear.
    \item Music ControlNet \citep{Wu2023MusicCM}: Music ControlNet is a training-based approach that shares the same underlying base model as our work but additionally fine-tunes adaptor modules during large scale training to the control signal $\bm{y}$ as conditioning.
    \item DOODL \citep{Wallace2023EndtoEndDL}: DOODL\footnote{\href{https://github.com/salesforce/DOODL}{\texttt{https://github.com/salesforce/DOODL}}} is an optimization-based approach that uses the EDICT \citep{wallace2023edict} sampler and multiple ad-hoc changes to the optimization process such as injecting noise and renormalizing $\bm{x}_T$. We use the same learning rate as \acro{} due to similar stability issues. 
\end{itemize}

We compare with Naïve Masking, MultiDiffusion, and Guidance Gradients for inpainting, outpainting, and looping experiments since they all have linear feature matching objective, Music ControlNet for the melody and intensity experiments, and FreeDoM and DOODL for all experiments.

\section{Results} 

\begin{figure}
    \centering
    \includegraphics[width=0.47\textwidth]{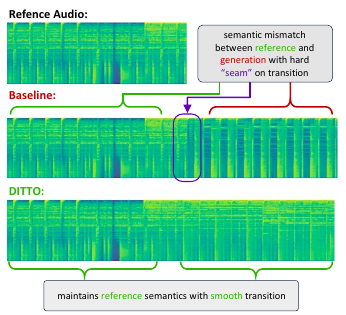}
    \caption{Failure cases of baseline outpainting methods. Baseline methods tend to create audible ``seams'' in the audio between overlap and non-overlap regions of the generated output, leading to unnatural jumps in semantic content. \acro{} avoids this issue and provides seamless outpainting throughout the full generation.}
    \label{fig:out}
\end{figure}

\subsection{Outpainting, Inpainting, and Looping  Results}\label{sec:paintres}

We show objective evaluation results for outpainting and looping in~\tref{tab:outpainting} and inpainting results in~\tref{tab:inpainting}. Here we report FAD, as low loss over the overlap regions does not necessitate that the \emph{overall} audio is cohesive. We find \acro{} achieves the lowest FAD against all baselines across overlap sizes of 1 to 3 seconds and inpainting gaps of 2 to 4 seconds. DOODL performs next behind \acro{}, and the inference-time guidance methods particularly struggle. 

Qualitatively, we discover that all baselines (besides DOODL) tend to produce audible ``seams'' in the output music outside the overlap region as shown in~\fref{fig:out}, wherein the final outputs tend to purely match the overlap region (i.e. over optimizing for the feature matching target) and ignore the overall consistency between the overlap generation and the rest of the generation. By optimizing $\boldsymbol{x}_T$ for reconstruction over the overlap regions, 
\acro{} 
effectively avoids such issues, as this process implicitly encourages the non-overlap generation sections to preserve semantic content seamlessly.

\begin{table}
    \footnotesize
    \centering
    \caption{Outpainting and looping FAD $(\downarrow)$ results for \acro{} against baseline pixel, guidance, and optimization-based methods.}
    \begin{tabular}{l|cccc}
    \toprule
         Method &  $o=1$&  $o=2$&  $o=3$& Looping\\
         \midrule
         DOODL& 0.719& 0.707& 0.700& 0.750\\
         Naive&  0.722&  0.716&  0.712& 0.753\\
         MD&  0.733&  0.716&  0.710& 0.749\\
         MD-50&  0.718&  0.714&  0.705& 0.752\\
         GG& 0.754& 0.738& 0.719& 0.774\\
         FreeDoM&  0.726&  0.723&  0.715& 0.758\\
         DITTO (ours)&  \textbf{0.716}&  \textbf{0.703}&  \textbf{0.698}& \textbf{0.746}\\
         \bottomrule
    \end{tabular}\label{tab:outpainting}
\end{table}

\begin{table}
\footnotesize
    \centering
    \caption{Inpainting FAD $(\downarrow)$ results for \acro{} against baseline pixel, guidance, and optimization-based methods.}
    \begin{tabular}{l|ccc}
    \toprule
         Method &  gap = 2&  gap = 3& gap = 4\\
         \midrule
         DOODL& 0.688& 0.693& 0.696\\
         Naive&  0.697&  0.705& 0.707\\
         MD&  0.690&  0.694& 0.701\\
         MD-50&  0.701&  0.708& 0.711\\
         GG&  0.700&  0.709& 0.717\\
         FreeDoM&  0.704&  0.709& 0.719\\
         DITTO (ours)&  \textbf{0.686}&  \textbf{0.688}& \textbf{0.690}\\
         \bottomrule
    \end{tabular}
    \label{tab:inpainting}
\end{table}

\subsection{Intensity, Melody, and Structure Results}

\begin{table*}[t!]
    \centering
    \footnotesize
    \caption{Intensity, melody, and structure control results. \acro{} achieves SOTA intensity and melody control. Music ControlNet struggles on intensity control MSE. FreeDoM performs well on structure but struggles on more complex melody and intensity control.}
    \label{tab:ctrl_res}
    \begin{tabular}{l|ccc|ccc|ccc}
    \toprule
         Control&  \multicolumn{3}{c|}{\textbf{Intensity}}  & \multicolumn{3}{c|}{\textbf{Melody}} &  \multicolumn{3}{c}{\textbf{Structure}}\\
         Metric&  MSE $(\downarrow)$&  FAD $(\downarrow)$&  CLAP $(\uparrow)$&  Acc $(\uparrow)$&  FAD $(\downarrow)$&  CLAP $(\uparrow)$ & MSE $(\downarrow)$& FAD $(\downarrow)$&CLAP $(\uparrow)$\\
    \midrule
         Default TTM&  40.843&  0.707&  0.373&   10.527 &  0.707&  0.373 & 0.309& 0.707&0.373\\
        ControlNet & 38.411& \textbf{0.637}& 0.308& 81.353& \textbf{0.545}&  \textbf{0.478} & --& --&--\\
 FreeDoM& 23.292& \underline{0.673}& \textbf{0.482}& 31.544& 0.706& \underline{0.477}& \textbf{0.018}& 0.668&\underline{0.415}\\
        DOODL & \underline{4.785} &0.695& 0.342 & \underline{81.592}& 0.715& 0.336 & 0.074 & \underline{0.653}& 0.387 \\
         DITTO (ours) &  \textbf{4.758}&  0.682&  \underline{0.433}&  \textbf{82.625}&  \underline{0.699}&  0.432 & \underline{0.024} & \textbf{0.632}&\textbf{0.418}\\
        \bottomrule
    \end{tabular}
\end{table*}

In~\tref{tab:ctrl_res}, we show objective metrics for intensity, melody, and structure control. We seek to understand 1) how different methods impose the target control on the generative model via MSE or Accuracy 2) overall audio quality via FAD and 3) how such control effects the baseline text conditioning via CLAP. We find \acro{} achieves SOTA intensity and melody control, beating that of Music ControlNet with \emph{zero} supervised training. 
We further explore Music ControlNet's poor 
intensity control more in-depth in Appendix~\ref{sec:corr_rms}. 
Additionally, we note FreeDoM slightly beats \acro{} in structure control, but exhibits poor performance for intensity and especially melody control, showing the limits of guidance-based methods 
for complicated feature extractors.

A notable concern with optimization-based control is the chance of reward hacking~\citep{skalse2022defining, Prabhudesai2023AligningTD}, where the control target is over-optimized leading to degradation in model quality and base behavior. We find that DOODL exhibits this reward hacking behavior consistently in addition to generally being worse at control than \acro{}, sacrificing overall quality and significant text relevance in favor of matching the control target. \acro{}, on the other hand, is able to balance the target control without over-optimizing and maintain quality and text relevance. 

In~\fref{fig:contr}, we show qualitative intensity, melody, and structure control results. On the left, we show a generated spectrogram with a rising then falling intensity curve. In the middle, we show a generated spectrogram with an input target and generated melody visualization (chromagram). On the right, we show a generated spectrogram with target and generated self-similarity matrices with an ABBA structure pattern.

\subsection{Subjective Listening Test}\label{sec:subj}

Given that audio quality is subjective, we performed a small scale listening test to measure the efficacy of \acro{} against alternative methods. Specifically, we asked test participants to rate the audio quality for three different applications including Intensity, Outpainting, and Melody across several algorithms. We generated 10 random samples for each applications using the same text prompts and control for each method. We compare \acro{} with FreeDoM and Music ControlNet for Intensity and Melody control, and with FreeDoM and MD-50 for outpainting. For each triplet of outputs for the given controls, participants were asked to rate the overall quality of the generated music for each output on a 0-100 scale. We recruited 15 participants for the listening study, thus totaling 150 scores per setting and control method.

\begin{table}
    \centering
    \footnotesize
    \caption{Subjective listening test results. \acro{} is strongly preferred to FreeDoM and Music ControlNet / MD-50 on outpainting and intensity tasks, and is roughly equivalent to Music ControlNet on melody control.}
    \label{tab:subjlist}
    \begin{tabular}{lcc}
    \toprule
 \multicolumn{3}{c}{\textbf{Intensity}}\\
           Comparison Test&  \% Wins & Avg. Difference\\
    \midrule
           \acro{} vs. ControlNet&  65 &\textbf{15.90} ($\pm$ 2.79)\\
           \acro{} vs. FreeDoM&  71 &\textbf{20.35} ($\pm 2.55)$ \\
    \midrule
  \multicolumn{3}{c}{\textbf{Outpainting}}\\
 Comparison Test& \% Wins & Avg. Difference\\
        \midrule
           \acro{} vs. MD-50&  77 &\textbf{23.49} ($\pm$ 2.57) \\
           \acro{} vs. FreeDoM&  80 &\textbf{26.17} ($\pm$ 2.33)\\
        \midrule
  \multicolumn{3}{c}{\textbf{Melody}}\\
 Comparison Test& \% Wins & Avg. Difference\\
        \midrule
           \acro{} vs. ControlNet&  48 &1.40 ($\pm$ 2.28)\\
           \acro{} vs. FreeDoM&  61 &\textbf{9.55} ($\pm$ 2.05)\\
    \bottomrule
    \end{tabular}
\end{table}

In Table~\ref{tab:subjlist}, we show the number of wins for \acro{} and the average difference in rating scores between \acro{} and each other method (where positive score difference denotes \acro{} is higher). Notably, we find that \acro{} is strongly preferred against FreeDoM on all tasks, Music ControlNet on Intensity, and MD-50 on Outpainting. On Melody control, we find practically no difference between \acro{} and Music-ControlNet, with \acro{}'s winrate at 48\% but having a slightly higher average score when favored. This provides evidence that \acro{} has superior or equal quality over SOTA controllable music generation methods.

\subsection{Efficiency Comparison}\label{sec:perfcomp}

Besides comparing \acro{} with DOODL in terms of their generation quality and control, we seek to understand how they differ in terms of both practical efficiency and convergence speed, as slow per-iteration runtime could be offset by fast convergence, and how such behaviors change as the number of sampling steps increases.
We focus on intensity control since it represents a middle ground between the simple linear painting methods and the more complex melody control. 
Besides MSE, FAD, and CLAP, we also report the mean steps to convergence (MS2C), i.e.~the average number of optimization steps needed to reach an MSE below some threshold $\tau$, the mean optimization speed (MOS), i.e.~the average number of seconds per optimization step, and the mean allocated memory (MAM), measuring the average GPU memory (in GB) used during optimization by the diffusion model. See Appendix~\ref{sec:effdetails} for more details.

In Table~\ref{tab:perf_abl}, we empirically confirm that DOODL is $\approx2$x slower than \acro{} and takes up $\approx2$x more GPU memory,
as DOODL
uses the EDICT sampler which doubles the number of model calls during both the forward and checkpointed backwards pass and stores both chains of inputs in memory. Most saliently, we discover that
DOODL displays practically identical convergence speed to \acro{}, showing that DOODL's added complexity provides no benefit in speeding up optimization.
We note that increasing the number of sampling steps tends to degrade control adherence, likely since the longer sampling chain makes backpropagation more difficult.
Interestingly, as sampling time increases the overall FAD improves significantly for DOODL, giving evidence that EDICT particularly struggles with few sampling steps, and thus DOODL cannot be sped up by using fewer steps without noticeable reward hacking.

We note that inference-time optimization-based techniques are slower than both guidance-based techniques and training-based techniques at inference time by design, as they functionally amortize the cost of the training-based methods (which require hundreds of GPU hours to fine-tune) at inference-time to offer more expressivity than the guidance-based methods (see Appendix~\ref{sec:effdetails} for more discussion). Given that the speed of \acro{} is primarily tied to the number of sampling steps used to sample the model (as well as the need for gradient checkpointing), there are clear ways to accelerate \acro{} using the growing line of work in fast diffusion samplers \cite{Lu2022DPMSolverFS, Luo2023LatentCM, Kim2023ConsistencyTM}, which we leave for future work.

\begin{table}
    \centering
    \footnotesize
    \caption{Performance between \acro{} and DOODL on intensity control. \acro{} and DOODL reach convergence in a similar number of steps yet DOODL is $\approx$2x less efficient than \acro{}. }
    \label{tab:perf_abl}
    \begin{tabular}{l|cc|cc}
    \toprule
         Method&  \acro{}&  DOODL&  \acro{} & DOODL\\
         Sampling Steps &  20&  20&  50& 50\\
         \midrule
         MSE $(\downarrow)$ &  \textbf{4.758}&  4.785&  \textbf{7.640}& 8.894\\
         FAD $(\downarrow)$ &  \textbf{0.682}&  0.695&  0.661& \textbf{0.636}\\
         CLAP $(\uparrow)$& \textbf{0.433}&  0.342&  \textbf{0.398}& 0.311\\
         MS2C $(\downarrow)$&  \textbf{44.466}&  49.203&  \textbf{46.855}& 47.834\\
         MOS$(\downarrow)$&  \textbf{1.859}&  4.177&  \textbf{4.472}& 10.036\\
         MAM $(\downarrow)$& \textbf{5.002} & 8.274 & \textbf{5.094}& 8.311\\
         \bottomrule
    \end{tabular}
\end{table}

\subsection{The Expressive Power of the Diffusion Latent Space}
Typically, the initial latent $\bm{x}_T$ is ignored in diffusion models, as the diffusion latent space has previously been
thought to encode little semantic meaning compared to GAN latent spaces \citep{song2020denoising, preechakul2022diffusion}.
\acro{}'s strong performance, however,
presents the surprising fact that a wide-array of semantically meaningful fine-grained features can be manipulated purely through exploring
the diffusion latent space 
without ever editing the pre-trained diffusion base model.
We explore this idea further, and how our findings are theoretically tied to the encoding of low-frequency structure noted by~\citet{Si2023FreeUFL} in Appendix~\ref{sec:lowfreq}.

\section{Conclusion}
We propose \acro{}: \acrolong{}, a unified training-free framework for controlling pre-trained diffusion models to enable a wide-range of creative editing and control tasks for music generation. \acro{} achieves SOTA editing ability and matches the controllability of fully training-based methods, outperforms the leading optimization-based approach while being  2x as time and memory efficient, and
imposes no restrictions on the modeling architecture or sampling process. 
In future work, we hope to accelerate the optimization procedure to achieve real-time interaction and more expressive control.

\section*{Impact Statement}
While generative multimedia models may open up new avenues for artistic creation, there is the concern of negatively impacting current working musicians and creators and their own livelihoods. We find that it is exceedingly important to build TTM systems that protect artists and their data. To mitigate harm, we train on licensed music and place our focus on improving controllability, allowing working artists to interface with TTM systems through more musically-aligned controls, instead of only relying on high-level textual prompts that may be too general for music professionals.

\bibliography{example_paper}
\bibliographystyle{icml2024}

\newpage
\appendix
\onecolumn

\section{Diffusion Review}\label{appendix:diffusion}
Denoising~diffusion~probabilistic~models~(DDPMs)~\citep{sohl2015deep,ho2020denoising} or diffusion models are a class of generative latent variable model. They are defined by a forward and reverse random Markov process.  Intuitively, the forward process takes clean data and iteratively corrupts it with noise to train a (denoising) neural network and the reverse process takes random noise and iteratively refines it with the learned network to generate new data. 

The forward process is defined as a Markov chain:
\begin{align}
    q(\bm{x}_{0},...,\bm{x}_{T}) &\defeq q(\bm{x}_{0}) \prod_{t=1}^T q(\bm{x}_{t} | \bm{x}_{t-1}) \label{eq:forwardmarkov}\\
    q(\bm{x}_{t} | \bm{x}_{t-1}) &\defeq \mathcal{N}(\sqrt{1 - \beta_t}\bm{x}_{t-1}, \beta_t \bm{I}) \label{eq:forwardsampling}
\end{align}
where $q(\bm{x}_{0})$ is the true data distribution, $q(\bm{x}_{T})$ is a standard normal Gaussian distribution, \mbox{$0 <~\beta_1 < \beta_2 < \dots < \beta_T$} are noise schedule parameters, and $T$ is the total number of noise steps.  To improve the efficiency of the fixed forward data corruption process, ~\eref{eq:forwardsampling} can be simplified to
\begin{align}
      q(\bm{x}_{t} | \bm{x}_{0}) &\defeq \mathcal{N}(\sqrt{\bar{\alpha}_t}\bm{x}_{0}, (1-\bar{\alpha}_t) \bm{I}) \label{eq:onestepforwardsampling}\\
      \bm{x}_{t} &\defeq \sqrt{\bar{\alpha}_t}\bm{x}_{0}+ \sqrt{1-\bar{\alpha}_t}\bm{\epsilon} \, , \label{eqn:add-noise}
\end{align}
where $\alpha_t = 1-\beta_t$,~$\bar{\alpha}_t = \prod^t_{i=1} \alpha_t$, and $\bm{\epsilon}$ is standard normal Gaussian noise, enabling forward sampling for any step $t$ given clean data $\bm{x}_{0}$. 

Given the forward process, we can specify a model distribution $p_\theta(\bm{x}_{0})$ that approximates $q_\theta(\bm{x}_{0})$. To make $p_\theta(\bm{x}_{0})$ easy to sample from, we specify the data generation process to be a \begin{align}
    p_\theta(\bm{x}_{0}) &= \int p_\theta(\bm{x}_{0},...,\bm{x}_{T})d\bm{x}_{1,...,T}\\
p_\theta(\bm{x}_{0},...,\bm{x}_{T})  &\defeq p_\theta(\bm{x}_{T}) \prod _{t=1}^{T} p_\theta^{(t)}(\bm{x}_{t-1} | \bm{x})\label{eq:reversemarkov}
\end{align}
where $\bm{x}_{0},...,\bm{x}_{T}$ are latent variables all in same data space.

Given the true data generation process~\eref{eq:forwardmarkov} and model~\eref{eq:reversemarkov}, we can train a neural network to recover the intermediate noisy data $\bm{x}_{t-1}$ given $\bm{x}_{t}$. More specifically, Ho et al.~\cite{ho2020denoising} showed that if we optimize the variational lower bound \citep{kingma2013auto} of our data likelihood and we reparameterize our problem to predict the noise $\bm{\epsilon}$, we can learn a suitable neural network $\bm{\epsilon}_\theta(\bm{x}_{t}, t)$ with parameters $\theta$ via minimizing the mean squared error via:
\begin{equation}
    \mathbb{E}_{\bm{x}_{0}, \bm{\epsilon}, t}\Big[ \lVert \bm{\epsilon} - \bm{\epsilon}_\theta(\bm{x}_{t}, t) \rVert^2_2 \Big] \, , \label{eqn:objective}
\end{equation}
where $t$ is the diffusion time-step. 

Given a learned $\bm{\epsilon}_\theta(\bm{x}_{t}, t)$, we can generate new data via the reverse diffusion process, a.k.a.~sampling.  To do so, we sample random Gaussian noise $\bm{x}_{T} \sim \mathcal{N}(0,I)$ and then iteratively refine it via
\begin{equation}
    \bm{x}_{t-1} = \frac{1}{\sqrt{\alpha_t}} \Big ( \bm{x}_{t} - \frac{1-\alpha_t}{\sqrt{1-\bar{\alpha}_t} } \bm{\epsilon}_\theta(\bm{x}_{t}, t) \Big) + \sigma_t  \bm{\epsilon}, \label{eq:DDPM_app}
\end{equation}
until $t=0$ to create our generated data $\bm{x}_{0}$ after $T$ denoising iterations. To obtain high-quality generations, $T$ is typically large (e.g., $1000$), which results in a slow generation process.

To reduce the computational cost of sampling (inference),~\citet{song2020denoising} proposed denoising diffusion implicit models (DDIM). DDIM uses an alternative variation optimization objective that itself yields an alternative sampling formulation 
\begin{align}
  &\begin{aligned}
    \bm{x}_{t-1} = 
    \sqrt{\alpha_{t-1}} \Bigg ( \frac{\bm{x}_{t} - \sqrt{1-\alpha_t}\bm{\epsilon}_\theta(\bm{x}_{t}, t)}{\sqrt{\alpha_t}}   \Bigg ) \\
    + \sqrt{1-\alpha_{t-1} - \sigma_t^2} \bm{\epsilon}_\theta(\bm{x}_{t}, t)
    + \sigma_t  \bm{\epsilon}, \label{eq:DDIM} 
    \end{aligned}
\end{align}
where $\bm{\epsilon} \sim \mathcal{N}(0,I)$, $\alpha_0 \defeq 1$, and $\sigma_t$ and different random noise scales. This formulation minimizes the number of sampling steps needed during inference (e.g., $50\sim100$) with minimal impact on generation quality.  Furthermore,
special cases of DDIM are then two fold 
1) when $\sigma_t = \sqrt{(1-\alpha_{t-1})/(1-\alpha_t)}\sqrt{1-\alpha_t/\alpha_{t-1}}$, DDIM sampling refers back to basic DDPM sampling and 
2) when $\sigma_t = 0$ the sampling process becomes fully deterministic.

To improve text conditioning, classifier-free guidance (CFG) can be used to blend conditional and unconditional generation outputs and trade-off conditioning strength, mode coverage, and sample quality~\cite{ho2022classifier}. When training a model with CFG, conditioning 
is randomly set to a null value a fraction of the time. During inference, the diffusion model output $\bm{\epsilon}_\theta(\bm{x}_{t}, t, \bm{c}_\text{text})$ is replaced with 
\begin{equation}
    \hat{\bm{\epsilon}}_{CFG} = w \cdot \bm{\epsilon}_\theta(\bm{x}_{t}, t, \bm{c}_\text{text}) + (1-w) \cdot  \bm{\epsilon}_\theta(\bm{x}_{t}, t, \bm{c}_\emptyset),
\end{equation}
where $\bm{c}_\text{text}$ are  text embeddings, $w$ is the CFG scaling factor, and $\bm{c}_\emptyset$ are null embeddings.

\section{EDICT and DOODL with invertible layers}\label{sec:edict_doodl}

Exact Diffusion Inversion via Coupled Transformations, or EDICT, is a sampling method introduced in \citet{wallace2023edict} to enable \emph{exact} diffusion inversion. EDICT accomplishes this by denoising two correlated diffusion chains, $\bm{x}_t'$ and $\bm{x}_t''$, at once, with the following updates:
\begin{align*}
    \bm{x}_t'^{\text{inter}} &= \sqrt{\frac{\alpha_{t-1}}{\alpha_t}}\bm{x}_t' + \left(\sqrt{1-\alpha_{t-1}} - \sqrt{\frac{\alpha_{t-1}(1-\alpha_t)}{\alpha_t}}\right)\boldsymbol{\epsilon}_\theta(\bm{x}_t'', t)\\
    \bm{x}_t''^{\text{inter}} &= \sqrt{\frac{\alpha_{t-1}}{\alpha_t}}\bm{x}_t'' + \left(\sqrt{1-\alpha_{t-1}} - \sqrt{\frac{\alpha_{t-1}(1-\alpha_t)}{\alpha_t}}\right)\boldsymbol{\epsilon}_\theta(\bm{x}_t'^{\text{inter}}, t)\\
    \bm{x}_{t-1}' &= p\bm{x}_t'^{\text{inter}} + (1-p)\bm{x}_t''^{\text{inter}}\\
    \bm{x}_{t-1}'' &= p\bm{x}_t''^{\text{inter}} + (1-p)\bm{x}_{t-1}',\\
\end{align*}
where the first two lines denote affine coupling layers and the last two lines are mixing layers with a fixed mixing coefficient $p$. This sampling procedure has the benefit of being exactly invertible:
\begin{align*}
    \bm{x}_{t+1}''^{\text{inter}} &= \frac{\bm{x}_t'' - (1-p)\bm{x}_{t}'}{p}\\
    \bm{x}_{t+1}'^{\text{inter}} &= \frac{\bm{x}_t' - (1-p)\bm{x}_{t+1}''^{\text{inter}}}{p}\\
    \bm{x}_{t+1}'' &= \sqrt{\frac{\alpha_{t+1}}{\alpha_{t}}}\left(\bm{x}_{t+1}''^{\text{inter}} - \left(\sqrt{1-\alpha_{t}} - \sqrt{\frac{\alpha_{t}(1-\alpha_{t+1})}{\alpha_{t+1}}}\right)\boldsymbol{\epsilon}_\theta(\bm{x}_{t+1}'^{\text{inter}}, t+1)\right)\\
    \bm{x}_{t+1}' &= \sqrt{\frac{\alpha_{t+1}}{\alpha_{t}}}\left(\bm{x}_{t+1}'^{\text{inter}} - \left(\sqrt{1-\alpha_{t}} - \sqrt{\frac{\alpha_{t}(1-\alpha_{t+1})}{\alpha_{t+1}}}\right)\boldsymbol{\epsilon}_\theta(\bm{x}_{t+1}'', t+1)\right)
\end{align*}
One consequence of the dual-chain sampling approach is the inherent tradeoff in setting the $p$ mixing parameter, as $p$ needs to be sufficiently low to prevent the two chains from diverging (especially at low sampling steps), and sufficiently high to prevent numerical precision errors when inverting the chains.

\begin{figure}[h]
    \centering
    \includegraphics[width=\textwidth]{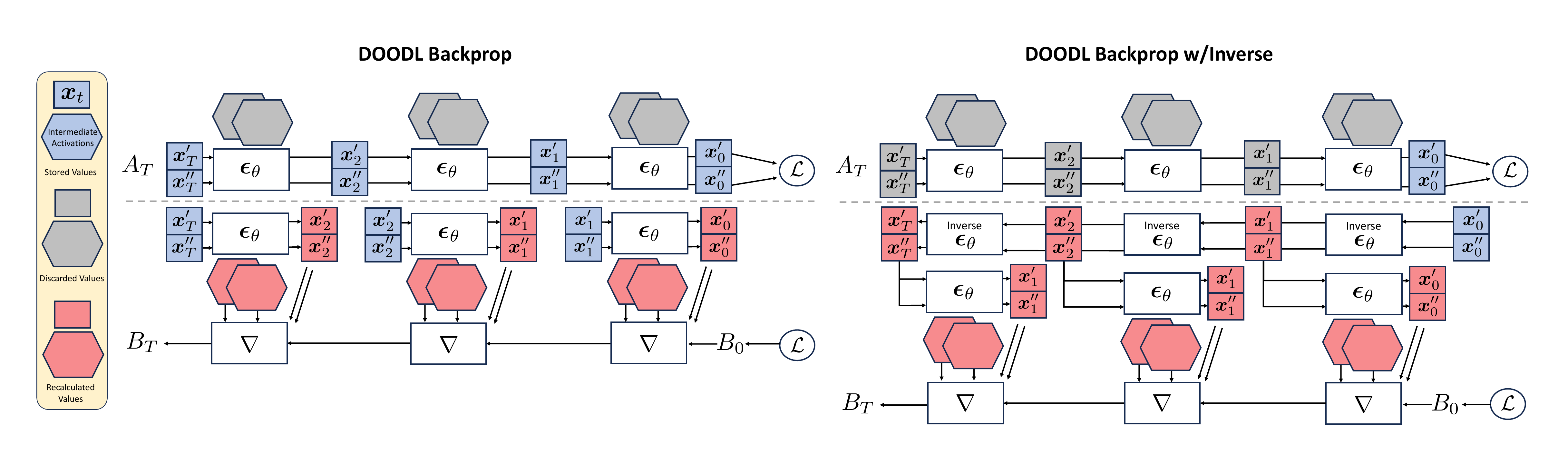}
    \caption{Forward and Backward pass for DOODL, both in its official implementation and alternatively by using the EDICT invertible layers. The standard DOODL backprop doubles the number of model calls (relative to \acro{}) due to the EDICT sampling, yet uses checkpointing to store function inputs for each timestep. When utilizing EDICT's invertibility, only the final outputs are stored in memory, yet the inversion process requires two \emph{more} model passes per timestep during the backwards pass.}
    \label{fig:doodl_inv}
\end{figure}

In the official implementation for DOODL, EDICT's invertibility is not used, and instead normal checkpointing is used on the EDICT sampler, thus using 4x the number of model calls as standard backpropagation. However, given the invertible nature of EDICT, DOODL can alternatively be formulated to directly use the inverse operation rather than storing \emph{all} function inputs in memory. In this setup, only the final $\bm{x}_0$ is stored in GPU memory, and then the inverse sampling operation is used to recalculate the function inputs, which are then passed back through the model to recalculate the intermediate activations for gradient calculation. This procedure is more memory efficient than the official implementation of DOODL and \acro{}, yet \emph{sextuples} the number of model calls and runtime, thus being the slowest procedure for inference-time latent optimization. Figure~\ref{fig:doodl_inv} describes both setups more in detail.

\section{Correlation-Based Intensity Control}\label{sec:corr_rms}

Given the surprising poor control performance of Music ControlNet \citep{Wu2023MusicCM} on the intensity control task despite being fully trained on such inputs, we investigated alternative metrics for understanding control adherence. Notably, we find that Music ControlNet implicitly models the intensity \emph{correlation}, paying more attention to the overall shape of the intensity curve across time than the absolute dB values of the curve itself. We believe this makes sense, given the UNet backbone convolution (correlation) layers are both scale and location invariant. Given this result, we can alternatively parameterize intensity control to directly optimize for correlation by setting $\mathcal{L} \propto -\rho(f(\bm{x}_0), \bm{y})$, or by maximizing the correlation between the target and output intensity curves. 

\begin{table}[h]
    \centering
        \caption{Intensity correlation results for Music ControlNet and \acro{} with both the standard and correlation-based loss function. By optimizing for correlation instead of absolute intensity, we can match the correlation of Music ControlNet while improving audio quality and text relevance.} 
    \label{tab:rms_corr}
    \begin{tabular}{l|cccc}
    \toprule
         Method&  MSE $(\downarrow)$&  $\rho$ $(\uparrow)$&  FAD $(\downarrow)$& CLAP $(\uparrow)$\\
         \midrule
         Music ControlNet&  \underline{38.4108}&  \textbf{0.9413}&  11.1315& 0.3084\\
         DITTO ($\mathcal{L} \propto ||f(\bm{x}_0)- \bm{y}||_2^2$)&  \textbf{4.7576}&  0.6166&  \textbf{10.5294}& \textbf{0.4326}\\
         DITTO ($\mathcal{L} \propto -\rho(f(\bm{x}_0), \bm{y})$)&  60.8952&  \underline{0.9040}&  \underline{11.0858}& \underline{0.3503}\\
         \bottomrule
    \end{tabular}
\end{table}

In Table~\ref{tab:rms_corr}, we show both the absolute MSE and correlation $\rho$ values for Music ControlNet, \acro{}, and \acro{} with the correlation based loss function. Music ControlNet has exceptional performance for intensity correlation, while baseline \acro{} unsurprisingly prioritizes absolute intensity over correlation given its optimization objective. By switching to the correlation objective, \acro{} can nearly match the correlation performance of Music ControlNet, all the while maintaining some of the absolute intensity \acro{}'s performance in audio quality and text relevance. This experiment shows how a single target feature can be parameterized in \acro{}'s flexible setup in multiple ways to change the intended behavior for rapid experimentation.

\section{\acro{} for Real-Audio Inversion}\label{sec:inversion}

Inversion, or the task of encoding real reference media $\boldsymbol{x}_{\text{ref}}$ into a generative model's latent space, is crucial for image and audio editing tasks~\citep{song2020denoising,dhariwal2021diffusion,Xia2021GANIA, mokady2023null}. Past audio-domain inversion work is very limited while past image-domain  methods include naively adding noise to inputs \citep{song2020denoising}, reversing the DDIM sampling process~\citep{dhariwal2021diffusion}, and learning additional \emph{null-text} parameters to improve inversion accuracy \citep{mokady2023null}. We use \acro{} for the task of inversion by setting $f(\bm{x}_0) = \bm{x}_0$, $\bm{y}=\bm{x}_{\text{ref}}$, and the loss to be the MSE or $\mathcal{L} \propto ||f(\bm{x}_0)-\bm{y} ||_2^2$. Then, we can solve~\eref{eq:problem} to find an $\boldsymbol{x}_T$ such that~\eref{eq:sampler} will produce $\boldsymbol{x}_0$ that reconstructs the target reference media $\boldsymbol{x}_{\text{ref}}$. While high-quality reconstruction is trivially possible with the fully invertible EDICT sampler \citep{wallace2023edict}, further editing with inverted content is complicated by its dual chain fully-deterministic sampling~\citep{pan2023effective}.

For generative text-conditioned models, a key factor of the inversion equation is the scale of the \emph{classifier-free guidance} parameter, which helps improve controllability through text \citep{ho2022classifier}, but noticeably makes the inversion process more difficult, as using classifier-free guidance results in diverging from the simple DDIM-based inversion \citep{mokady2023null}. Against \acro{}, we compare with the Naïve inversion method of simply adding Gaussian noise to the reference spectrogram, the DDIM-based inversion which runs the DDIM sampling process in reverse through the model, and the recent Null-Text Inversion \citep{mokady2023null} method, which starts with the DDIM inversion and then learns a time-dependent unconditional text embedding $\bm{c}_{\emptyset, t}$ to improve inversion results in the presence of high guidance scales. Like in null-text, we use the DDIM inversion as an initial guess for \acro{}. 

As the goal is direct recreation of the reference audio, we report MSE reconstruction across the entire 5K-sample MusicCaps dataset. 
We run this evaluation across four different guidance scales (ranging from 0, which is purely unconditional, to 7.5), and additionally run this on both our baseline 6 second model as well as a \emph{24} second music generation model, which maintains all the same training hyperparameters and model size as our base model and only differs in that the output dimension is $2048\times160\times1$. In Table~\ref{tab:inv_res}, we show that \acro{} beats all other inversion methods across all guidance scales and model sizes, with the exception of the highest guidance scale on the 6 second base model, for which it performs slightly worse than null-text inversion. Notably, \acro{}'s superior performance on the 24 second model shows that scaling the number of free parameters with the image size (as $\bm{x}_T$ is the same shape as the output spectrogram) helps maintain reconstruction quality in the presence of high guidance, while methods that do not scale with the image size (like null-text inversion) do not have this benefit. 

Qualitatively, we find that null-text inversion exhibits unique semantic artifacts in the reconstructed audio, such as replacing sung vocals with trumpets or tambourines with hi-hats, while \acro{} avoids this failure case. As all the training data for the base model was on purely instrumental music, this shows that \acro{} allows TTM diffusion models to interact with real audio outside the distribution of their training data. In further work, we hope to explore more complicated edits that require inverted inputs (which is common in the image domain) and thus compare against the EDICT-based approach.
\begin{table}[h]
    \centering
    \caption{Inversion results across context size and guidance strength. \acro{} performs SOTA reconstruction in most cases and noticeably scales with context size.}
    \label{tab:inv_res}
    \begin{tabular}{l|cccc|cccc}
    \toprule
\multirow{2}{*}{ MSE $(\downarrow)$}& \multicolumn{4}{c|}{6 seconds}& \multicolumn{4}{c}{24 seconds}\\
          &  $w = 0$&  $w = 1$&  $w = 4$& $w = 7.5$ & $w = 0$& $w = 1$& $w = 4$&$w = 7.5$ \\
         \midrule
         Naïve&  0.0678&  0.0668&  0.0714& 0.0787 & 0.1044& 0.1042& 0.1071&0.1122 \\
         DDIM&  0.0115&  0.0072&  0.0192& 0.0334 & 0.0089& 0.0072& 0.0115&0.0179 \\
         NT&  0.0043&  0.0072&  0.0055& \textbf{0.0072} & 0.0057& 0.0072& 0.0057&0.0060 \\
         DITTO (ours)&  \textbf{0.0011}&  \textbf{0.0010}&  \textbf{0.0025}& 0.0075 & \textbf{0.0011}& \textbf{0.0011}& \textbf{0.0015}&\textbf{0.0023} \\
    \bottomrule
    \end{tabular}
\end{table}

\section{Reference-Free Looping}\label{sec:norefloop}

\begin{figure}[h]
    \centering
    \includegraphics[width=0.7\textwidth]{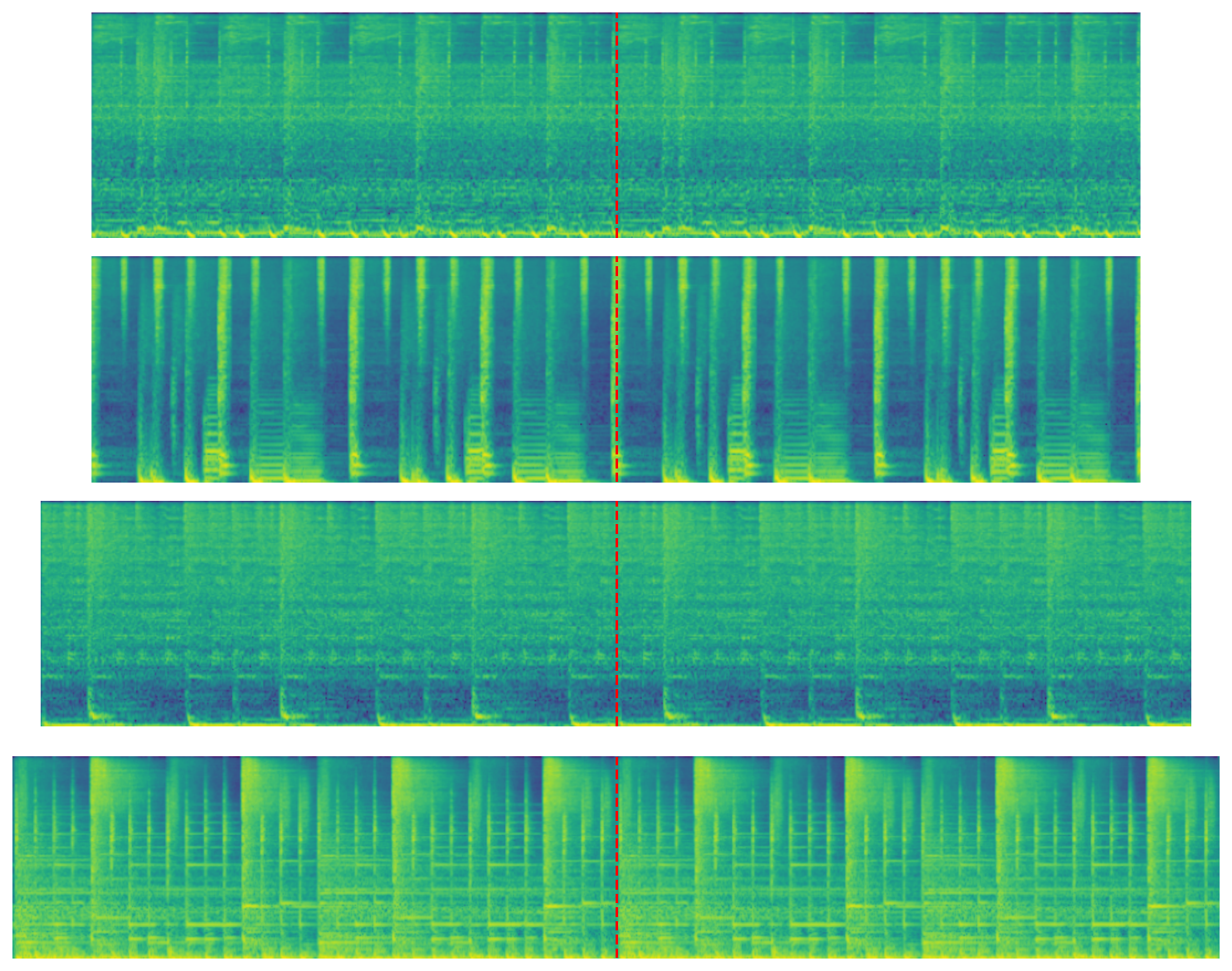}
    \caption{Reference-free loop generation with an overlap of $o=0.5$ seconds. Loop boundary is shown in red.}
    \label{fig:ref_free_loop}
\end{figure}

While we generally focus on long-form reference-based loop generation, where we seamlessly take existing audio and blend it back into itself, we note that \acro{} can also be used for short-form reference-\emph{free} loop generation, where we seek to generate a short  musical loop unconditionally. This framework is similar to the reference-based looping, but instead defines the generated audio to loop back into \emph{itself}, rather than into some fixed reference audio. More formally, we define $\mathbf{M}_{\text{gen}, 1}$ and $\mathbf{M}_{\text{gen}, 2}$ as two $o$ sized masks over the generated spectrogram, and set $f(\bm{x}_0) = \mathbf{M}_{\text{gen}, 1} \odot \bm{x}_0$, $\bm{y} = \mathbf{M}_{\text{gen}, 2} \odot \bm{x}_0$, and $\mathcal{L} \propto \|f(\bm{x}_0) - \bm{y}\|_2^2$, such that the model optimizes to match the overlap region of its own generation during \acro{}. We note that by setting $\mathbf{M}_{\text{gen}, 2}$ to occur earlier in the spectrogram (rather than one of the edges), we can generate loops of lengths that are less than or equal to the total context window (in our case, 6 seconds). In Figure~\ref{fig:ref_free_loop}, we show spectrograms of reference-free looping with an $o=0.5$ second overlap and a total of two repetitions, with the loop boundary shown in red.

\section{Musical Structure Transfer}\label{sec:fgstruct}

\begin{figure}
    \centering
    \includegraphics[width=0.6\textwidth]{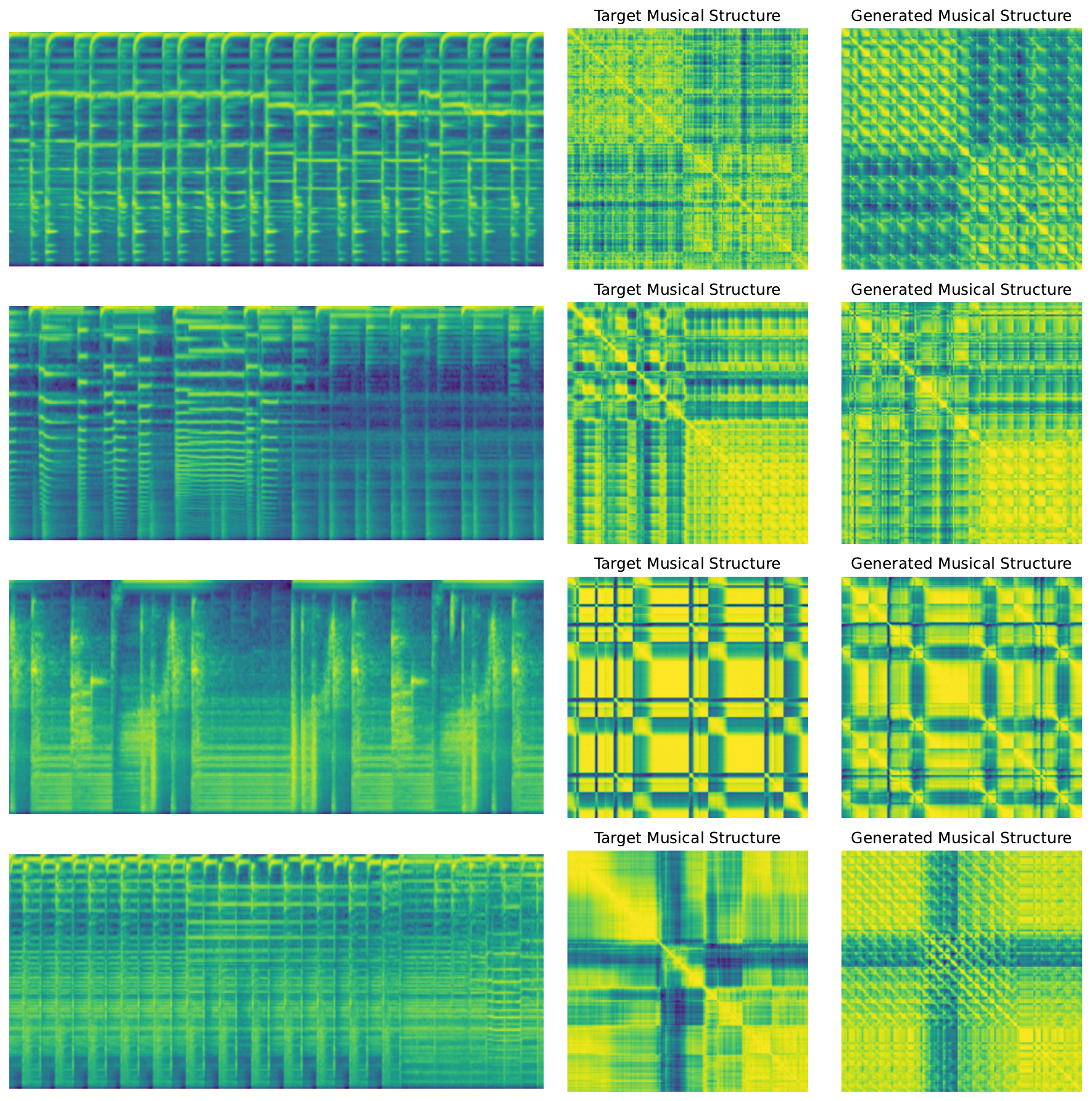}
    \caption{Musical Structure Transfer using self-similarity MFCC matrices extracted from real musical audio as the target.}
    \label{fig:fgstruct}
\end{figure}

While in the main paper, we focus our musical structure control task as controlling high-level musical form through simple musical phrase diagrams (like ``ABA"), we can also directly \emph{transfer} the structure of an existing song to our generation with \acro{} through a similar process. Namely, instead of generating a target self-similarity matrix based on a given phrase diagram, we can instead set $\bm{y} = \mathbf{T}(y)\mathbf{T}(y)^\top$, where $y$ is the mel-spectrogram of a \emph{real} song and $\mathbf{T}(\cdot)$ is our MFCC-based timbre-extraction function. In this way, using $\mathcal{L} \propto \|f(\bm{x}_0) - \bm{y}\|_2^2$ we can use \acro{} to generate music that matches the fine-grained self-similarity matrix of an existing musical fragment. Note that here we omit the 2D Savitzky-Golay step over the output self-similarity matrix, as here we want to directly match the intra-phrase similarity structures (rather than trying to capture broad musical form). We show examples of spectrograms with the target and generated self-similarity matrices in~\fref{fig:fgstruct}, where target self-similarity matrices are extracted from songs from the Free Music Archive dataset~\citep{fma_dataset}.

\section{Alternative Sampling Methods}\label{sec:alt_samp}

Unlike previous works on diffusion latent optimization \citep{Wallace2023EndtoEndDL}, \acro{} imposes no restrictions on the sampling process used to perform the optimization procedure, thus freeing us to choose any performant diffusion model sampling algorithm. Namely, we explore using DPM-Solver++~\citep{Lu2022DPMSolverFS}, a SOTA diffusion sampler for improving sample quality in conditional diffusion settings. Using outpainting and intensity control as test cases, in Table~\ref{tab:dpm} we show MSE and FAD results. We interestingly find that DDIM is \emph{better} than DPM++ for the intensity control task, yet DPM++ is slightly better for the outpainting task. We invite future work on discovering both theoretically and empirically how different diffusion sampling algorithms effect the noise latent optimization process.

\begin{table}[h]
    \centering
    \caption{Comparison of different samplers for \acro{}. DDIM works solidly better than DPM++ for the intensity task, and DPM++ preforms slightly better for outpainting.}
    \label{tab:dpm}
    \begin{tabular}{cccc}
    \toprule
        Target & Sampler &  MSE & FAD\\
    \midrule
         Intensity &  DDIM & 4.77 & 10.53\\
         Intensity & DPM++ &  6.30 & 11.04\\
         Outpainting&  DDIM& -- & 9.19\\
         Outpainting&  DPM++& -- & 9.12\\
    \bottomrule
    \end{tabular}
    
\end{table}

\section{Multi-Objective \acro{}}\label{sec:multi_obj}

Inspired by~\cite{Wu2023MusicCM}, we can leverage the flexibility of \acro{} to incorporate \emph{multiple} feature matching criteria for a multi-objective optimization setup:
\begin{equation}\label{eq:moditto}
    \boldsymbol{x}_T^* = \arg \min_{\boldsymbol{x}_T} \frac{1}{M}\sum_{i=1}^M\lambda_i\mathcal{L}_i\left(f_i(\boldsymbol{x}_0), \boldsymbol{y}_i\right),\\
\end{equation}
where we include additional $\lambda_i$ weights to balance the different scales of each loss function. Given \acro{}'s generality, this allows us to combine both editing \emph{and} control signals at the same time, effectively unlocking the ability to iteratively compose long-form music with fine-grained temporal control. Here, we experiment with Intensity+Structure and Intensity+Melody, showing the combination of multiple reference-free controls, and Intensity+Outpainting, showing how reference-free controls can be composed with reference-based editing methods. For Intensity+Outpainting and Intensity+Structure we set $\lambda_{\text{intensity}}=1/40$ and set $\lambda_{\text{intensity}}=1/4$  for Intensity+Melody, while all other $\lambda_i=1$, as intensity is calculated in the raw dB space. For the Intensity+Outpainting control, we use an overlap of $o=2$ seconds and only optimize the intensity curve for the \emph{nonoverlapping} section, having a similar effect to the ``don't care'' regions in \citet{Wu2023MusicCM}. Here we compare against FreeDoM \cite{yu2023freedom} for all tasks and Music-ControlNet \cite{Wu2023MusicCM} for the Intensity+Melody task.

\begin{table*}[t!]
    \centering
    \caption{Multi-objective control results. \acro{} More effectively balances multiple control signals than FreeDoM and Music ControlNet.}
    \label{tab:multiobj}
    \begin{tabular}{l|ccc}
    \toprule
         Control&  \multicolumn{3}{c}{\textbf{Intensity+Outpainting}}  \\
         Method&  Intensity MSE ($\downarrow$)&  FAD ($\downarrow$)&  CLAP ($\uparrow$)\\
         \midrule
        DITTO& \textbf{5.783}& \textbf{0.699}& \textbf{0.506}\\
 FreeDoM& 23.945& 0.705& 0.502\\
          \midrule
         &  \multicolumn{3}{c}{\textbf{Intensity+Structure}}\\
 & Intensity MSE / Structure MSE ($\downarrow$)& FAD ($\downarrow$)& CLAP ($\uparrow$)\\
    \midrule
 DITTO& \textbf{6.802} / \textbf{0.092}& \textbf{0.661}& 0.432\\
 FreeDoM& 21.033 / 0.304& 0.669& \textbf{0.490}\\
          \midrule
 & \multicolumn{3}{c}{\textbf{Intensity+Melody}}\\
 & Intensity MSE ($\downarrow$) / Melody Acc ($\uparrow$)& FAD ($\downarrow$)& CLAP ($\uparrow$)\\
          \midrule
 DITTO& \textbf{7.833} / 0.436& 0.680& 0.405\\
 FreeDoM& 21.185 / 0.198& \textbf{0.683}& \textbf{0.494}\\
 Music-ControlNet& 37.841 / \textbf{0.452}& 0.604& 0.347\\
          \bottomrule
    \end{tabular}
\end{table*}

In Table~\ref{tab:multiobj}, we find that FreeDoM in general struggles to follow multiple control signals across most tasks, while \acro{} is able to more effectively balance the competing optimization objectives. Interestingly, we find generally low performance on the Intensity+Melody task across all methods, which leave for future work.

\begin{figure}
    \centering
    \includegraphics[width=.7\textwidth]{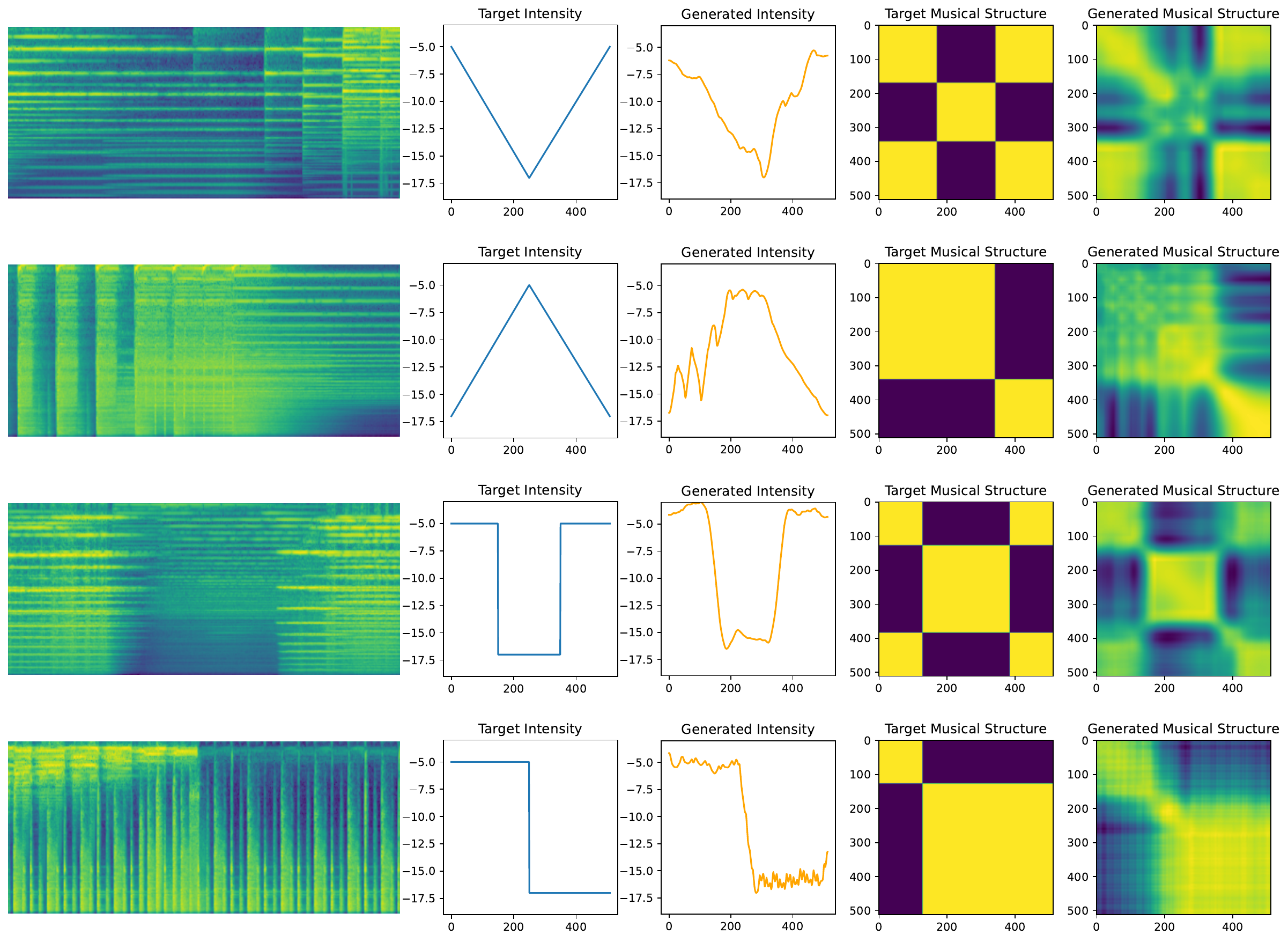}
    \caption{Output spectrograms, intensity curves, and MFCC self-similarity matrices for multi-objective DITTO with intensity and structure set as the feature extractors. }
    \label{fig:multiop_rmsmfcc}
\end{figure}

\begin{figure}
    \centering
    \includegraphics[width=0.7\textwidth]{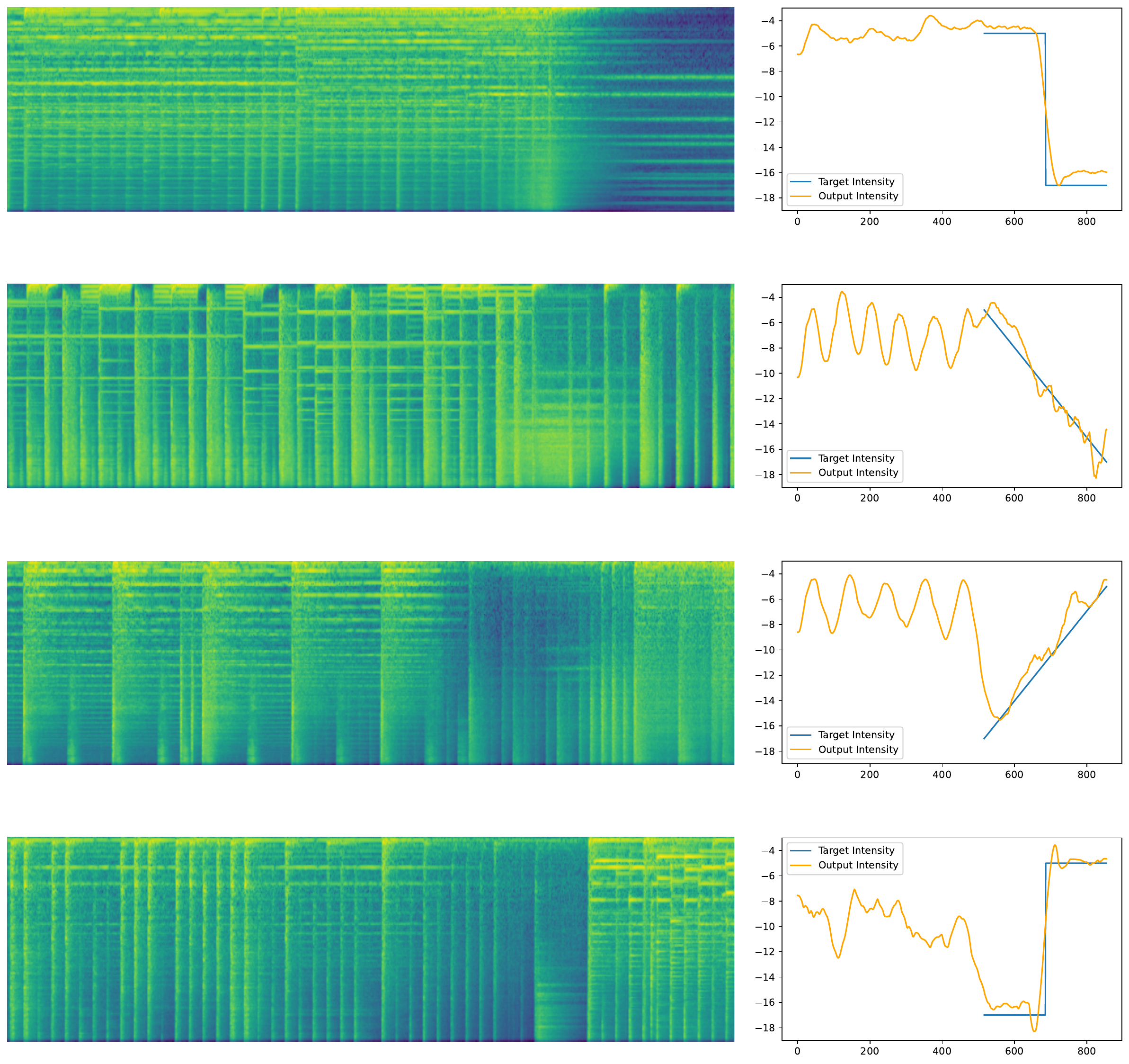}
    \caption{Output spectrograms and intensity curves for multi-objective DITTO with outpainting and intensity set as the feature extractors. The overlap is set to $o=2$ seconds, and intensity control is only applied over the non-overlapping section.}
    \label{fig:multiop_rmsout}
\end{figure}

In Figures~\ref{fig:multiop_rmsmfcc} and~\ref{fig:multiop_rmsout}, we show spectrograms and output features for both experiments.

\section{Reusing Optimized Latents}\label{sec:reuse}

A key bottleneck of inference-time optimization methods like \acro{} is the apparent need for the optimization procedure to generate a single output that matches the given feature, thus limiting its scalability. In order to mitigate this effect and accelerate the creative workflow for users, we explore how we can \emph{reuse} optimized latents $\bm{x}_T^*$ to generate diverse outputs that follow the initial optimized feature signal. 

A natural idea to add reusability to optimized latents is to treat each $\bm{x}_T^*$ as the mean of some normal distribution $\mathcal{N}(\bm{x}_T^*, \sigma^2)$ within the model's latent space for some hyperparameter $\sigma^2$, and then sample an $\bm{x}_T \sim \mathcal{N}(\bm{x}_T^*, \sigma^2)$ at inference time without re-optimizing. We find that this process leads to considerable divergence from the optimized feature in practice, and leave this to future work to explore further. Instead, we consider the case where we sample stochastic \emph{trajectories} starting from $\bm{x}^*_T$, which in practice is as simple as switching to a stochastic sampling algorithm at inference time such as DDPM in~\eref{eq:DDIM} (note that we still use \emph{deterministic} samplers during \acro{} as stochastic samplers tend to make the optimization process considerably harder). Additionally, we also explore the case when the initial prompt $\bm{c}_{\text{text}}$ used during \acro{} is varied, adding another source of stochasticity.  

In this experiment, we compare two possible methods for reusing optimized latents for sampling stochastic trajectories: 1) after performing \acro{} with DDIM, we sample using DDPM at inference time  and 2) we use DDIM for optimization and DDPM for inference, but then additionally include the FreeDoM~\citep{yu2023freedom} guidance update in each DDPM step. To test reusability, after optimizing for each $\bm{x}^*_T$ given a target signal $\bm{y}$ and some text condition $\bm{c}_{\text{text}}$, we generate $B$ samples $\bm{x}_0^{(i)}$ using $\bm{x}^*_T$ as the starting latent and our stochastic sampling algorithm of choice, and measure $\frac{1}{B}\sum_{i=1}^B \mathcal{L}(f(\bm{x}_0^{(i)}, \bm{y}))$, or the average loss over the stochastic samples, where \emph{no} optimization is occuring. We perform this experiment both where each $\bm{x}_0^{(i)}$ is generated with a random prompt $\bm{c}_i$, and when each prompt is fixed to the initial prompt $\bm{c}_i = \bm{c}_{\text{text}}$ to measure the effect of additional stochasticity from conditioning.

In Table~\ref{tab:reuse}, we show results for intensity, melody, and musical structure control with a batch size $B=10$. Notably, while switching to baseline DDPM during sampling predictably worsens the feature adherence, using FreeDoM with DDPM and starting at $\bm{x}_T^*$ yields significantly improved feature adherence to the optimized target. This presents a useful marriage of guidance-based and optimization-based approaches, as \acro{} latents can act as reasonable feature priors by utilizing FreeDoM to guide the trajectory from the strong starting point.

\begin{table}
    \centering
    \caption{Loss on samples generated with stochastic sampling from $\bm{x}^*_T$. We observe that \acro{} latents natively can act as generalized feature priors, using FreeDoM on optimized latents to significantly improve feature adherence, thus showing how optimization-based and guidance-based methods can be used in conjunction for high-quality and efficient control.}
    \label{tab:reuse}
    \begin{tabular}{lll|cc}
    \toprule
         Optimization&  Inference&  \multirow{2}{*}{Feature}& \multirow{2}{*}{$\mathcal{L}$} & $\mathcal{L}$\\
         \texttt{Sampler} & \texttt{Sampler} & & & (Fixed Prompt)\\
         \midrule
         DDIM&  DDPM&  Intensity& 24.5120& 13.8316\\
         DDIM&  DDPM+FreeDoM&  Intensity& 16.9780& 11.2481\\
         DDIM&  DDPM&  Melody& 2.7973 & 2.7441\\
         DDIM&  DDPM+FreeDoM&  Melody& 1.8482& 1.8710\\
         DDIM&  DDPM&  Musical Structure& 0.2952& 0.2643\\
         DDIM&  DDPM+FreeDoM&  Musical Structure& 0.0251& 0.0235\\
         \bottomrule
    \end{tabular}
    
\end{table}

\section{Diffusion Latents and Low-Frequency Content}\label{sec:lowfreq}

In~\citet{Si2023FreeUFL}, the authors discover that much of the low-frequency (in the 2D pixel domain) content of TTI model generations are determined exceedingly early on in the sampling process, where further sampling steps only produce high-frequency information and improve quality. This presents a compelling case for why \acro{} has such strong expressivity: because many target controls for TTM generation like intensity, melody and musical structure are low-frequency features in the spectrogram domain (i.e. most high-frequency 2D content in spectrograms address audio quality factors), optimizing $\bm{x}_T$ to target these features is well within the diffusion model's latent space which already encodes low-frequency information in the first place. This is compounded by the fact that music tags and captions generally only address high-level stylistic information, leaving everything that is not captured by the text captions (such as time-varying intensity, melody, and structure) to be incorporated into the initialization.

\begin{figure}
    \centering
    \includegraphics[width=\textwidth]{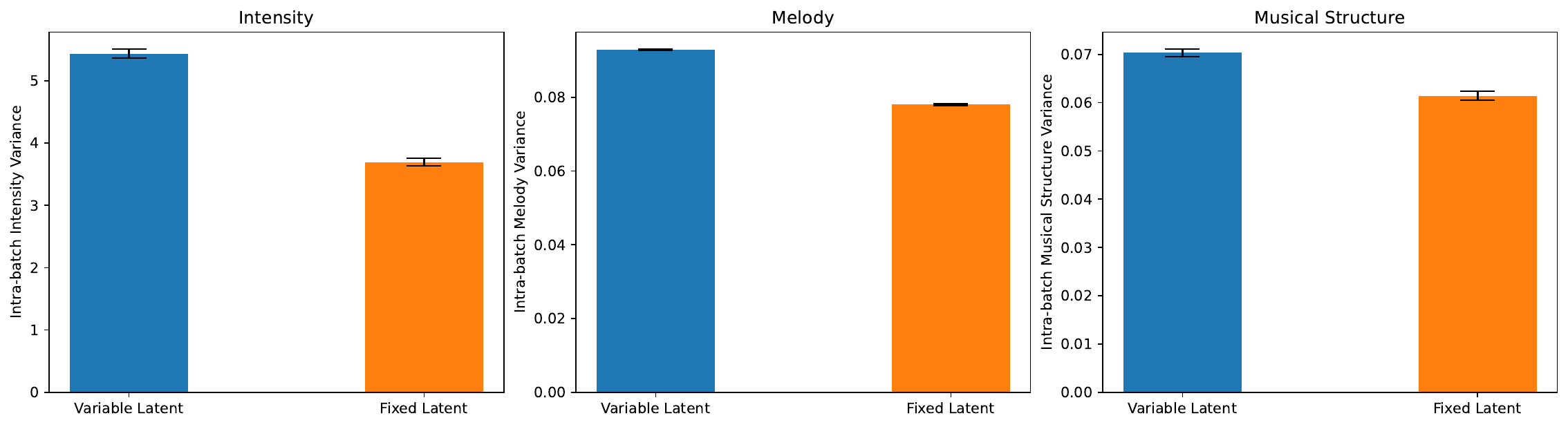}
    \caption{Intra-batch variance for model generations both with and without fixing the initial latent. We find a statistically significant effect that fixing the latent reduces feature variance, showing that $\bm{x}_T$ already encodes a great deal of feature information.}
    \label{fig:initstruct}
\end{figure}

To validate this proposed justification, we generate 5K batches ($B=10$) of samples from our base diffusion model, where half of the batches (2.5K) have random initializations and random prompts while the other half have the same initialization $\bm{x}_T$ (and still random prompts). For each group, we measure variance within each batch of the intensity, melody, and musical structure features extracted from the batch outputs. Shown in~\fref{fig:initstruct}, we find a statistically significant effect across all features that fixing the initialization significantly reduces the intra-batch feature variance. This serves as empirical justification that to a certain extent, the model output's salient musical features are already determined \emph{at initialization}.

\section{Model Pre-training}\label{sec:diff-model}
For our spectrogram generation model, we follow an identical training processed to default TTM as to Music ControlNet~\cite{Wu2023MusicCM}. We use a convolutional UNet~\cite{ronneberger2022convolutional} with 5 2D-convolution ResNet~\cite{he2016deep} blocks with $[64, 64, 128, 128, 256]$ feature channels per block with a stride of 2 in between downsampling blocks. The UNet inputs Mel-scaled~\cite{stevens1937scale} spectrograms clipped to a dynamic range of 160 dB and scaled to $[-1, 1]$  computed from 22.05 kHz audio with a hop size of 256 (i.e., frame rate $\mathrm{f_k} \approx 86$ Hz), a window size of 2048, and 160 Mel bins. For our genre, mood, and tempo global style control $\bm{c}_\text{text}$, we use learnable class-conditional embeddings with dimension of 256 that are injected into the inner two ResNet blocks of the U-Net via cross-attention. We use a cosine noise schedule with 1000 diffusion steps that are injected via sinusoidal embeddings with a learnable linear transformation summed directly with U-Net features in each block. We set our output time dimension to 512 or $\approx$6 seconds, yielding a 512$\times$160$\times$1 output dimension. We use an L1 training objective between predicted and actual added noise, an Adam optimizer with learning rate to $10^{-5}$ with linear warm-up and cosine decay. Due to limited data and efficiency considerations, we instantiate a relatively small model of 41M parameters and pre-train with distributed data parallel for 5 days on 32 A100 GPUs with a batch size of 24 per GPU.  Finally, we also use MusicHifi~\cite{Zhu2024MusicHiFiFH} as the vocoder: MusicHifi uses a BigVGAN vocoder~\cite{lee2022bigvgan} modified  with a DAC discriminator~\cite{kumar2023high}, trained with an AdamW optimizer with learning rate 0.0001, exponential learning rate decay on both our discriminator and generator optimizer, batch size of 48 per GPU, and 1536 channels for the initial upsampling layer that was trained on 8 A100 GPUs for 5 days.

\section{Efficiency Experiment Details and Discussion}\label{sec:effdetails}
We run the test on a single 40GB A100 with $K=70$ maximum optimization steps and $\tau = 2$ dB. For DOODL, we use a mixing coefficient of $p=0.93$ at 50 steps following \citet{Wallace2023EndtoEndDL} and $p=0.83$ at 20 steps due to severe divergence issues with higher $p$ at 20 steps. 

\begin{table}[h!]
    \centering
    \caption{Speed comparison of various training-based, guidance-based, and optimization-based methods on the intensity control task, both in fine-tuning cost (in 40GB A100 GPU hours) and latency.}
    \label{tab:effcomp}
    \begin{tabular}{l|cc}
    \toprule
         Method&  Fine-tuning Cost (GPU Hours, $\downarrow$)& Latency (seconds, $\downarrow$)\\
         \midrule
         Base TTM&  -& 0.612\\
         ControlNet \cite{Wu2023MusicCM}&  576& 1.456\\
         FreeDoM \cite{yu2023freedom}&  0& 2.867\\
         \acro{} (Ours) &  0& 82.192\\
         DOODL \cite{Wallace2023EndtoEndDL}&  0& 206.897\\
         \bottomrule
    \end{tabular}
\end{table}

To augment the analysis in Sec.~\ref{sec:effdetails} and display how \acro{}'s speed compares to other control methods for TTM diffusion models, in Table~\ref{tab:effcomp}, we report both the latency (i.e.~the time for a single sample to be generated, in seconds) and the overall fine-tuning cost in 40GB NVIDIA A100 GPU hours for our Base TTM model as well as \acro{}, DOODL, Music ControlNet, and FreeDoM. Notably, Music ControlNet presents the lowest inference latency at the cost of over 500 GPU hours of fine-tuning. Of the training-free methods, \acro{} is faster than DOODL but still $\approx$30x slower than FreeDoM, offering a clear trade-off in terms of latency and control strength.

\end{document}